\documentclass[%
 reprint,
 amsmath,amssymb,
 aps,
]{revtex4-2}
\usepackage{color}
\usepackage{graphicx}
\usepackage{dcolumn}
\usepackage{bm}
\usepackage{bbold}
\usepackage[font=scriptsize]{caption} 
\usepackage{subcaption}
\usepackage{amssymb}
\usepackage{amsmath}
\usepackage{xcolor}
\usepackage{bm} 
\usepackage{float} 

\newcommand{\me}{m_\mathrm{e}}
\newcommand{\smax}{s_\mathrm{max}}
\newcommand{\Lambdac}{\Lambda_\mathrm{crit}}

\begin{document}

\preprint{APS/123-QED}

\title{Ergodicity, lack thereof, and the performance of\\ reservoir computing with memristive networks}

\author{Valentina Baccetti}
\affiliation{%
School of Science, RMIT University, Melbourne, Victoria 3000, Australia
}%
\author{Ruomin Zhu}
\author{Zdenka Kuncic}
\affiliation{%
School of Physics, University of Sydney, Sydney, NSW 2006, Australia}%
\author{Francesco Caravelli}
\affiliation{%
 Theoretical Division (T4), Los Alamos National Laboratory, Los Alamos, New Mexico 87545, USA
}%

\date{\today}

\begin{abstract}
Networks composed of nanoscale memristive components, such as nanowire and nanoparticle networks, have recently received considerable attention because of their potential use as neuromorphic devices. In this study, we explore the connection between ergodicity in memristive and nanowire networks, showing that the performance of reservoir devices improves when these networks are tuned to operate at the edge between two global stability points. The lack of ergodicity is associated with the emergence of memory in the system. We measure the level of ergodicity using the Thirumalai-Mountain metric, and we show that in the absence of ergodicity, two memristive systems show improved performance when utilized as reservoir computers (RC). In particular, we highlight that it is also important to let the system synchronize to the input signal in order for the performance of the RC to exhibit improvements over the baseline.
\end{abstract}

\maketitle

\section{\label{sec:level1} Introduction}

Memristive networks are electronic circuits that use memristive devices, a type of resistive switching memory element. These networks store and recall information by modifying the device's resistance. They find applications in computer memory, neuromorphic computing, and signal processing \cite{Mehonic2020}. Memristive networks offer benefits such as high density, low power consumption, and potential for high-speed operation. Simultaneously, there is growing interest in alternative approaches to computation and optimization in response to the rapidly increasing demands on computing \cite{Oliver2019}. Various proposals have emerged to address this challenge, some of which involve the use of oscillators or frequency domain encoding \cite{Isingmachine,Pierangeli_2019,Vadlamani2020,Csaba2020,goto}, leveraging near- or in-memory computation \cite{Singh2019,Ielmini2018,DCRAM,Sebastian2020,traversa,Ventra2018}, and exploring memcomputing \cite{traversa,Ventra2018}. These innovative techniques aim to provide more efficient solutions for complex optimization problems \cite{Hennessy2019,Vadlamani2020,traversa,Sutton2017,Torrejon2017,Isingmachine,Pierangeli_2019, Kirkpatrick83,Dorigo2004}. However, understanding large assemblies of memristive devices in non-equilibrium statistical mechanics remains a challenge. Recent research has focused on exploring the geometric and statistical properties of nanowire networks, where electrical junctions exhibit memristive behavior due to the interplay between tunneling and filament formation phenomena\cite{Milano2019,Kuncic2021}.

Conductive nano-filament formation at nanowire-nanowire junctions creates a memristive device, while quantum tunneling contributes to additional nonlinearity in switching behavior, making the dynamics more complex \cite{Hochstetter2021}. As most of the voltage drop occurs at the junctions, a basic model for the conductance evolution of these networks involves an assembly of memristive devices with voltage or current generators. The dynamic behavior of these systems is currently being studied, especially regarding bias-induced conductance transitions
\cite{Diaz-Alvarez2019,caravelli2021,Zhu2021information,Hochstetter2021}.

Various conductance transitions have been observed in memristive devices, often characterized by transient unstable dynamics of the memristive components and their internal memory parameter. One well-known transition, which has been defined as ``rumbling", has been analytically identified in the simplest model of memristive networks \cite{caravelli2021}. A similar transition has also been observed in nanowire networks \cite{Hochstetter2021}. This transition involves the system effectively moving between different minima of an effective potential and is marked by bursts of transient positive Lyapunov exponents. It arises from the coexistence of multiple low-dimensional equilibrium points in the dynamics. The mean-field theory for such systems can be ensured by mapping them to a PEDS (projective embedding of dynamical systems) \cite{caravelli2022}. This is relevant to our study as we aim to identify non-ergodic behavior associated with these transitions.

In certain memristive systems, as the applied voltage increases, an effective mean potential develops with multiple minima. As one minimum becomes dominant, the system undergoes a rapid chaotic transition towards this emerging stable fixed point. This transition is distinct from the conventional Landau picture of symmetry breaking with bifurcation, as it arises from the competition between two minima.

While the understanding of these transitions is clearer in simplified memristive device models, it becomes less evident in more realistic systems.  In this paper, we aim to describe the dynamic behavior of these transitions in two systems using ergodicity measures. Ergodicity, a concept in statistical mechanics and thermodynamics, pertains to the long-term behavior of a system. It states that the time average of a system over an extended period is equivalent to the ensemble average of the system. In essence, it implies that the system reaches a steady state, and its long-term average behavior aligns with the average behavior across many instances of the same initial conditions.

Ergodicity plays a crucial role in various natural sciences, such as physics and chemistry, enabling an understanding of system evolution, equilibrium attainment, and the prediction of long-term behavior based on short-term observations. In the context of memory-dependent systems, the lack of ergodic behavior signifies (hard) ergodicity breaking, which occurs when symmetry breaking transpires in thermodynamic systems \cite{palmer}. 
Typically, physical systems operate within a regime where ergodicity holds for a subset of possible phase space states. Quantifying ergodicity and gaining insight into the dynamical state of a physical system is vital for comprehending the system's operational regime. One of the goals of this paper will also be to test the hypothesis that computation near a transition point improves. 

To test this hypothesis, we will use reservoir computing (RC) as our computational model \cite{Jaeger2004}, which has been recently shown to be universal \cite{Grigoryeva2018}. It has been reported for instance that the ``edge of chaos" may be important for the performance of RC \cite{Carroll2020}, but it is important to stress that critical states have been reported both for biological neuronal networks in the brain \cite{Tagliazucchi2012} and in other artificial neural networks \cite{Morales2021}.

This article is organized as follows: in section~\ref{sec:ergodicitymemoryTM} we introduce the general definition of ergodicity and how it is related to memory; in section~\ref{sec:memristivedevices} we introduce the two memristive systems we are considering in this study; in section~\ref{sec:RCandergodicity} we give the definition of the TM metric for the two memristive systems we are considering, and test the RC task results for both memristive systems in terms of ergodicity breaking. Conclusions follow.

\section{Memory, ergodic Convergence and the Thirumalai-Mountain metric}
\label{sec:ergodicitymemoryTM}

Ergodicity and the emergence of memory are closely intertwined concepts within the realms of statistical physics and complexity science. Ergodicity refers to a system's property of uniformly exploring all possible states in its phase space over an extended period. In essence, it characterizes a system's behavior as it traverses its accessible states in a time-averaged manner. On the other hand, memory denotes the influence of past states on the current state of a system.

In many systems, the emergence of memory is closely connected to the violation of ergodicity. When a system is ergodic, its behavior remains independent of its past history, exhibiting no memory effect. However, when ergodicity is violated, the system may display persistent or long-term correlations, resulting in the emergence of memory.

A notable example of the relationship between ergodicity and memory can be observed in spin-glasses, 
which are disordered magnetic systems that exemplify complex dynamics with persistent or long-term correlations that foster memory emergence. The behavior of spin glasses is often described through two-time correlation functions, which measure correlations between spin configurations at different points in time.



Studying two-time correlation functions in spin glasses sheds light on system dynamics, including the emergence of memory and ageing phenomena. These functions measure correlations between spin configurations at different times, while the auto-correlation function reveals details about the system's relaxation dynamics. Slow decays observed in these functions signify the violation of ergodicity and the emergence of memory, while the dependence on both $t$ and the waiting time $t_w$ highlights the ageing phenomenon in spin glasses \cite{Vincent2007} and other frustrated systems \cite{Saccone2022,saccone}.
Ergodicity refers to a system's property where the long-term behavior can be deduced from a single, extended observation of the system, rather than relying on multiple independent realizations.

According to Boltzmann's hypothesis, the trajectories of any dynamical system in its phase space eventually evolve into regions where macroscopic properties reach thermodynamic equilibrium \cite{dorfmann99a}. The ergodic hypothesis states that ensemble averages and time averages coincide as time progresses. This means that the ensemble-averaged value of an observable, denoted as $\langle g \rangle$, can be obtained by averaging its values over time during the observable's evolution. Mathematically for the observable $g(t)$
we have:

\begin{eqnarray}
\label{eq:erEn}
\langle g \rangle = \lim_{T \to \infty} \frac{1}{T}\int_{{0}}^{T} g(t) dt.
\end{eqnarray}

It is important to note that various definitions of ergodicity exist in the Physics and Mathematics literature \cite{mastatistical, mountain89me, ergodictheory}, with significant distinctions. For example, in Markov chains, ergodic behavior requires a strongly connected transition graph, which is rarely the case. In practice, a system can exhibit ergodicity within a specific subset of its phase space. The concept of ``effective ergodicity'' was introduced to describe scenarios where the system rapidly samples coarse-grained regions \cite{mountain89me}.

A measure of effective ergodic convergence is based on the observation that certain components of a system exhibit identical average characteristics at thermal equilibrium. These characteristics are determined by an observable defined on the system's phase space, denoted as $\Gamma$. To assess the effective ergodic behavior of the observable, it is necessary to estimate its average value using an ensemble approach, such as a thermal ensemble. This estimation is typically performed using the Thirumalai-Mountain (TM) $g$-fluctuating metric, denoted as $\Omega_g(t)$.

The TM $g$-fluctuating metric, introduced by Thirumalai and Mountain \cite{mountain89me, mountain_measures_1989, mountain93me}, quantifies the difference between the ensemble-averaged value of the observable, $g(\Gamma)$, and the sum of the instantaneous values of $g(t)$ for each component of the system. At a given time $t$, the TM $g$-fluctuating metric is expressed as:

\begin{eqnarray}
\label{eq:Og}
\Omega_{g}(t) = \frac{1}{N} \sum_{j=1}^{N} \big[ \bar g_{j}(t) - \langle g(t) \rangle \big]^{2},
\end{eqnarray}

Here, $\bar g_{j}(t)$ represents the time-averaged value per component, and $\langle g(t) \rangle$ denotes the instantaneous ensemble average, defined as:

\begin{eqnarray}
\label{eq:gEns}
\bar g_{j}(t) = \frac{1}{t} \sum_{i=0}^{t} g_{j}(t_i), \
\langle g(t) \rangle = \frac{1}{N} \sum_{j=1}^{N} g_{j}(t).
\end{eqnarray}

This definition assumes that $g(\Gamma)$ serves as a suitable physical order parameter, effectively characterizing the system's behavior. The definition of instantaneous ensemble average will depend on the considered system, as it will be shown in  Sec.~\ref{sec:RCandergodicity}.

The rate of ergodic convergence can be quantified using the derivative of the effective ergodic convergence, denoted as $\Omega_{g}^{'}$. This quantity is given by:

\begin{eqnarray}
\label{eq:Dg1}
\Omega_{g}^{'} = \frac{\Omega_{G}(t)}{\Omega_{G}(0)} \to \frac{1}{t D_{g}},
\end{eqnarray}
in the diffusive regime. When the system is instead in a sub-diffusive or super-diffusive regime, the power of the relaxation in time changes from 1 to a different exponent. 
Coarse graining of the phase space leads to the clustering of the system's accessible states, making the concept of effective ergodicity more applicable. Effective ergodicity is achieved when the system uniformly explores the coarse-grained regions within a finite time \cite{mountain_measures_1989}. Here, $D_{G}$ represents the diffusion coefficient associated with the property being studied, and $\Omega_{g}$ refers to the effective ergodic convergence. This definition aligns with other notions of ergodicity in cases where diffusion follows a power law \cite{barkai}.
The rate of ergodic convergence, determined using the TM metric, provides an estimate of the system's ergodicity. The behavior of the rate, as described by Eq.~\eqref{eq:Dg1}, indicates the system's attainment of effective ergodicity. For example, if the inverse of the rate scales linearly with time, the system reaches ergodicity in a diffusive manner: $1/{\Omega}_g \rightarrow D_G t$, where $D_G = \Omega_{G}(0)$ represents the diffusion coefficient of the property $G$. It is generally expected that the rate scales with time as $\Omega'_{G}(t) \sim t^{-p}$.

When the inverse of $\Omega_{g}^{'}$ exhibits a linear relationship with time, it indicates that all points in the phase space are equally likely, resembling the behavior of Brownian motion. This approach has been applied in various contexts, such as simple liquids \cite{mountain89me}, earthquake fault networks \cite{tiampo03a, tiampo2007a}, and the Ising model \cite{suzen}.

\section{Models of memristive devices}
\label{sec:memristivedevices}

We would like to introduce the notion of memristive device that we will use in the following. We will consider two models of memristive device, a resistive current-controlled memristive device, which can be described by the equations
\begin{eqnarray}
    V(t)&=&R\big(x\big) I(t),\\
    \frac{dx}{dt}&=&f(I,x). \nonumber
\end{eqnarray}
The second model we consider is a conductance based device, of the form
\begin{eqnarray}
    I(t)&=&G\big(\lambda\big)V(t),\\
    \frac{d\lambda}{dt}&=&f(V,\lambda).
\end{eqnarray}
Both models satisfy the pinched hysteresis property for the $I-V$ curve, characteristic of memristive devices~\cite{Chua1971}.

\subsection{Memristive network toy model}
\label{sec:memtoymodeltheory}
In \cite{Caravelli2017b}, the dynamical equation for a circuit of memristors was derived under the assumption of a resistance current-controlled device. The resistance function used, $R(x)=R_{on} (1-x)+x R_{off}$, approximates $TiO_2$ memristors, where $R_{on}$ and $R_{off}$ represent the limiting resistances, and $x\in[0,1]$ is \textit{internal memory parameter}, the state variable describing the size of the oxygen-deficient conducting layer. At the lowest order, the evolution of the internal memory parameter can be described by a simple equation with hard boundaries:

\begin{equation}
\frac{dx}{dt}=\frac{R_{off}}{\beta} I-\alpha x=\frac{R_{off}}{\beta} \frac{V}{R(x)}-\alpha x
\end{equation}

Here, $\alpha$ and $\beta$ are the decay constant and the effective activation voltage per unit of time, respectively, which determine the timescales of the system.

While the model presented above provides a simple description of a polar resistive device, various extensions have been explored in the literature. For instance, to account for diffusive effects near the boundaries, some models remove the hard boundaries and introduce a window function \cite{Joglekar2009,Prodromakis2011}. Although these models better capture the detailed IV curves of physical devices, they still exhibit the fundamental pinched hysteresis behavior observed in the linear model.


In adimensional units ($\tau = \alpha t$), the equation for $x(t)$ in a single memristor device under an applied voltage $S$ can be derived using Ohm's law ($S = RI$), \cite{Caravelli2016rl,caravelli2021}. The resulting equation is:

\begin{equation}
\frac{d}{d\tau}x = \frac{S}{\alpha \beta} \frac{1}{1 - \chi x} - x = -\partial_x V(x,s),
\label{eq:oned}
\end{equation}

Here, $\chi = \frac{R_{off}-R_{on}}{R_{off}}$ and $s = \frac{S}{\alpha \beta}$, with $0 \leq \chi \leq 1$ in relevant physical cases. The dynamics of the system, represented by a single memristor device, are fully characterized by the gradient descent in the effective potential given by \cite{caravelli2021}:

\begin{equation}
V(x,s) = \frac{1}{2}x^2 + \frac{s}{\chi}\log(1 - \chi x),
\label{eq:potential}
\end{equation}

\begin{figure}
    \centering
    \includegraphics[width=0.99\linewidth]{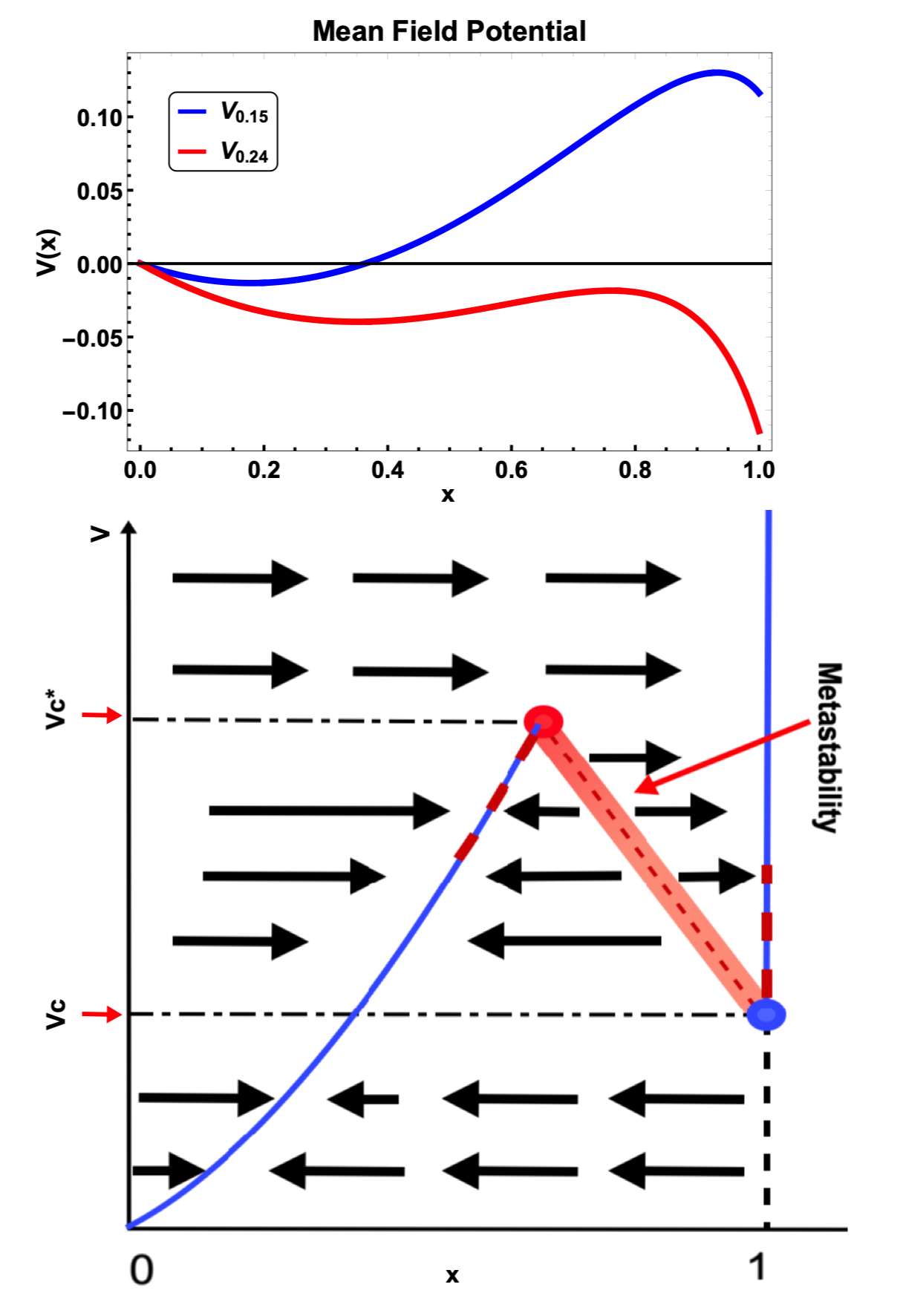}
    \caption{\textit{Top:} Evolution of the mean-field potential for the toy model as a function of voltage. \textit{Bottom:} Phase diagram of the toy model, and the appearance and disappearance of stable points. The intermediate region in voltage between $Vc$ and $Vc^*$ is characterized by the coexistence of two stable points. The buffer region corresponds to a metastable state which acts as the boundary between the attracting stable points for a single memristive device, while the shading represents the tunneling region for the network of devices.}
    \label{fig:mfp}
\end{figure}

The potential exhibits two minima separated by a barrier, as depicted in Fig. \ref{fig:mfp} (top), for $s = 0.15$ and $s = 0.24$ with $\chi = 0.9$. The range of $s$ for the existence of a barrier is limited, and when $\chi$ approaches 1, the local minimum can move inside the domain $[0,1]$, leading to the emergence of an unstable fixed point (i.e., the peak of the barrier). Consequently, two basins of attraction and locally stable minima are formed. A pictorial phase diagram illustrating this behavior is shown in Fig. \ref{fig:mfp} (bottom), highlighting the critical voltage points at which such behavior occurs. Below a critical voltage point $V_c$, only one fixed point is present. At the $V_c$ value, a new fixed point emerges, but it becomes metastable in the presence of noise (dashed red line). At an intermediate point $Vc < V_m < Vc^*$, the two metastable points have equal energy, representing the switching point where the original stable fixed point becomes metastable. At higher values of $V = Vc^*$, the first fixed point disappears by merging with the barrier and becoming flat. Ultimately, only one fixed point remains at higher values.

For a network of memristors of this type, the mean-field potential resembles exactly the single memristor defined above. For details, see Supp. Mat. \ref{sec:tmod}. The key difference is that the mean-field potential is only an approximation when the system is high dimensional, e.g. now there is an effective metastable region, as shown in Fig. \ref{fig:mfp}, instead of two stable minima, when one of the minima is lower than the other.  Such effective ``tunneling'' can be explained using the theory developed in \cite{caravelli2022}, i.e. the fact that local mean field maxima become saddle points in high dimensions. Here we will use this fact to study effective ergodicity breaking in memristive networks.

\begin{figure}
    \centering
\includegraphics[width=0.95\linewidth]{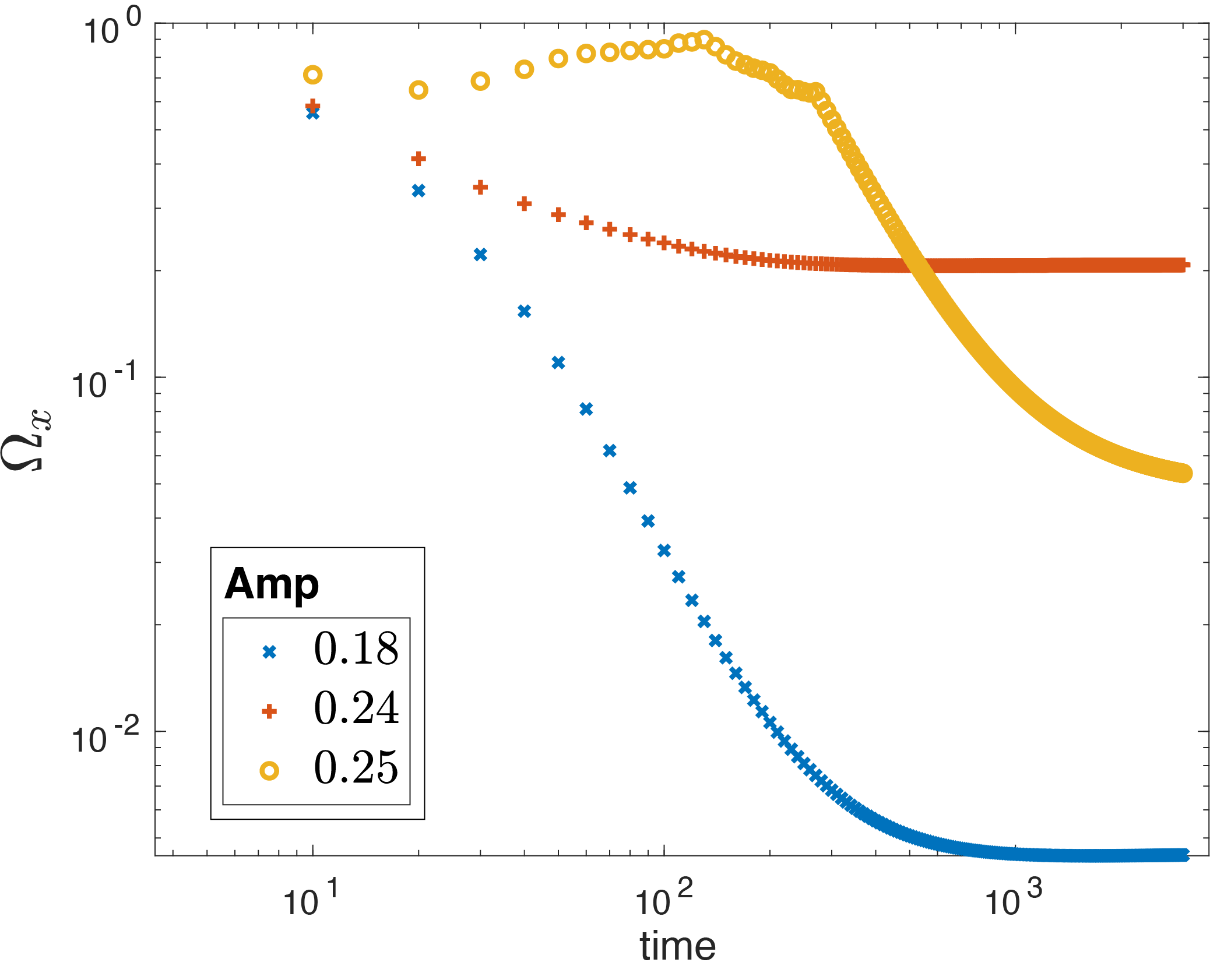}
    \caption{Thirumalai-Mountain metric as a function of time and amplitude, calculated for the stochastic memristive model with noise, with the sample average calculated using the mean field value introduced in~\eqref{eq:meanfield}. We observe that, in the regime in which the potential has a single minimum ($V<Vc$), the TM metric decays rapidly. Instead, for higher voltages in which two minima are present ($Vc<V<Vc^*$), the TM metric does not decay. This is a symptom of non-ergodic behavior.}
    \label{fig:tmw1}
\end{figure}

In fact, these two regimes can be characterized via the Thirumalai-Mountain (TM) metric introduced before. In Fig. \ref{fig:tmw1} we show the  TM metric as a function of time for $\bar s=0.18$ (top), $\bar s=0.24$ (center), and $\bar s=0.25$ (numerically calculated for a circuit of N = 50 memristors with parameters $\alpha = \beta = 1$ and $\chi = 0.9$). As we can see, in the former case the TM metric relaxes as a power law with an approximate intermediate exponent of $p\in[-2,-1.5]$, typical of a diffusive regime. For $\bar s=0.24$, we see instead a typical behavior of weak ergodicity breaking, e.g. a transient non-monotonicity of the TM metric, associated with the trajectories effectively ``tunneling'' through the mean field barrier. This behavior is indeed transient as for $\bar s = 0.25$ the TM metric relaxes again as a power law.

Using the toy model, we can then pinpoint such non-monotonicity of the TM metric to a transient transition between two stable asymptotic states and the effective symmetry breaking of the potential. Thus, we can immediately identify the symmetry breaking of the potential as the culprit of such transient non-ergodic behavior. 


\subsection{Nanowire networks}

We now consider a more realistic model of memristive networks, which can be associated with self-assemblies of silver nanowires.
Via established bottom-up self-assembly techniques,
one can readily synthesize nanowire networks (NWNs) \cite{Diaz-Alvarez2019,Li2020}. These NWNs typically have a 2D spatial distribution of randomly oriented nanowires that are interconnected by cross-point MIM junctions. The NWN we consider have densities of 10 junctions/$\mu$m2 and 0.5 nanowires/$\mu$m2. Device electrodes can be deposited onto the substrate using a mask, from which the conductance measurements can be readily performed.  This bio-inspired structure is difficult to design and fabricate using top-down techniques. 
 Self-assembled NWNs exhibit topological properties, such as small-world propensity and modularity, that are similar to biological neural networks and distinct from random and grid-like networks \cite{Loeffler2020}. Unlike fully connected bipartite networks in artificial neural networks, small-world networks have local connectivity and short path lengths, making them relatively sparse. Although small-worldness is necessary for important functional properties, such as synchronizability and information flow, it alone cannot explain the diverse range of dynamics across networks that exhibit this structural property.

\begin{figure}[h]
	\centering
    \includegraphics[width=1\linewidth]{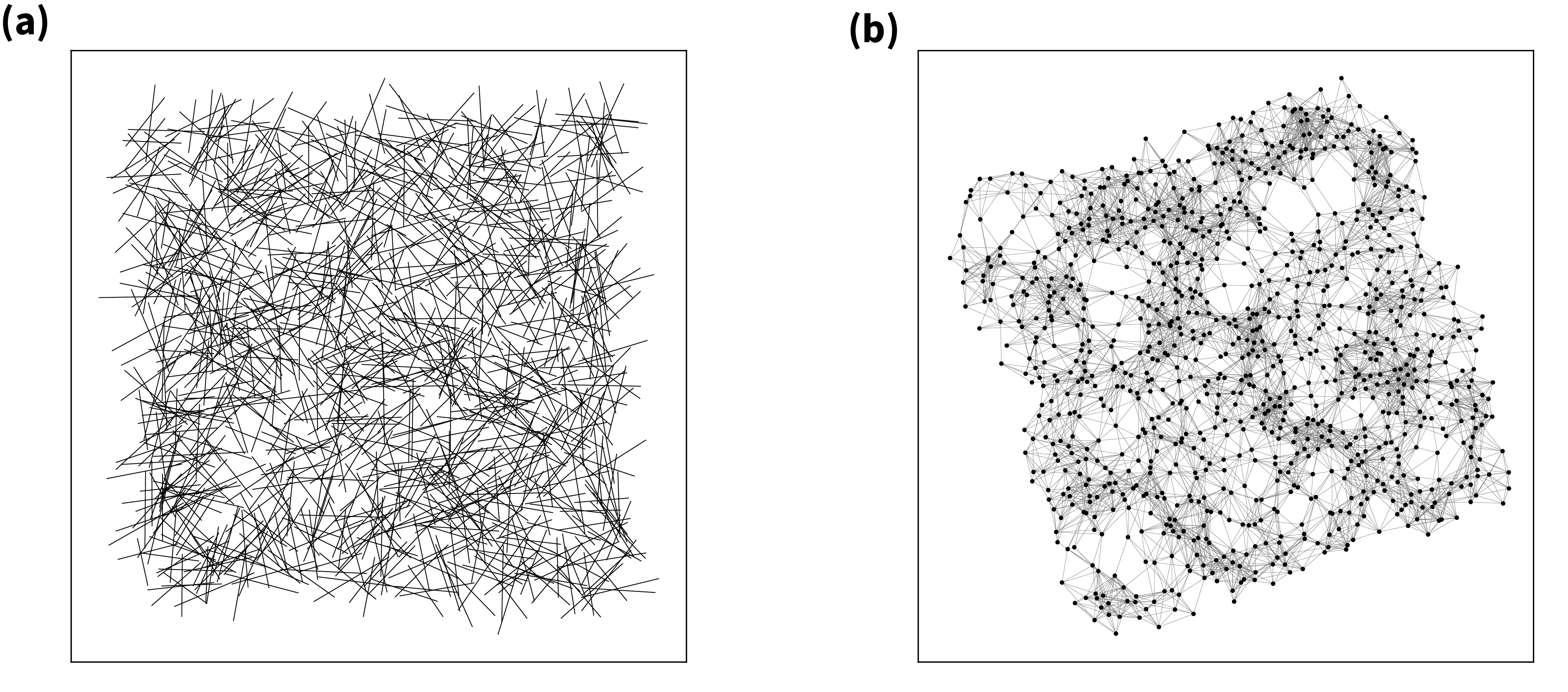}
	\caption{Example of a nanowire network generated with the random wire model. 
    (a) Simulated NWN with 1024 nanowires and 6877 junctions. (b) Graphical representation of the NWN in (a).
    }
	\label{fig:nws}
\end{figure}

A model for the simulation of realistic NWNs has been introduced previously in the literature \cite{Kuncic2020, Zhu2021information,Hochstetter2021}.
Fig.~\ref{fig:nws} (a) shows a visualization of a simulated nanowire network containing 1000 nanowires and 6877 junctions. Self-assembly is modeled by distributing nanowires on a $3 \times 3 \,\mu {\rm m}^2$ 2D plane, with their centers uniformly sampled from $[0,3]$ and orientation uniformly sampled from $[0, \pi]$. The lengths of the nanowires are sampled from a gamma distribution (mean = $100 \, $nm, standard deviation $10 \,$nm), based on experimental measurements \cite{Diaz-Alvarez2019}.
In theoretical studies, as illustrated in Fig.~\ref{fig:nws} (b), the NWN is transformed to the corresponding graphical representation, where the nodes represent nanowires and the edges are the cross-point junctions.

In this work, all simulation results for nanowire networks are generated using a network comprised of 1000 nanowires and 6877 junctions.
All variables, except the adjacency matrix $A$, are time-dependent.
A model for the conductance of a single junction, associated with the filament length is provided in the supplementary material, see Sec. \ref{sec:juncm}. Each junction evolves as voltage bias is continuously applied to the network. 
The modified nodal analysis approach is applied to the graphical representation to solve Kirchhoff's voltage and current conservation laws at each time step \cite{Nilsson}. This is equivalent to the method used for the derivation of the exact network equation for the toy model, eqn. (\ref{eq:manyd}). Although the NWN model is based on polymer-coated Ag nanowires, with memristive junction internal dynamics that differ from that of the toy model (based on metal-oxide memristors),
the network dynamics are similar and one should think of the two models as equivalent from a physical perspective. 

\begin{figure*}
    \centering
    \hspace{-0.1cm}\includegraphics[width=1\linewidth]{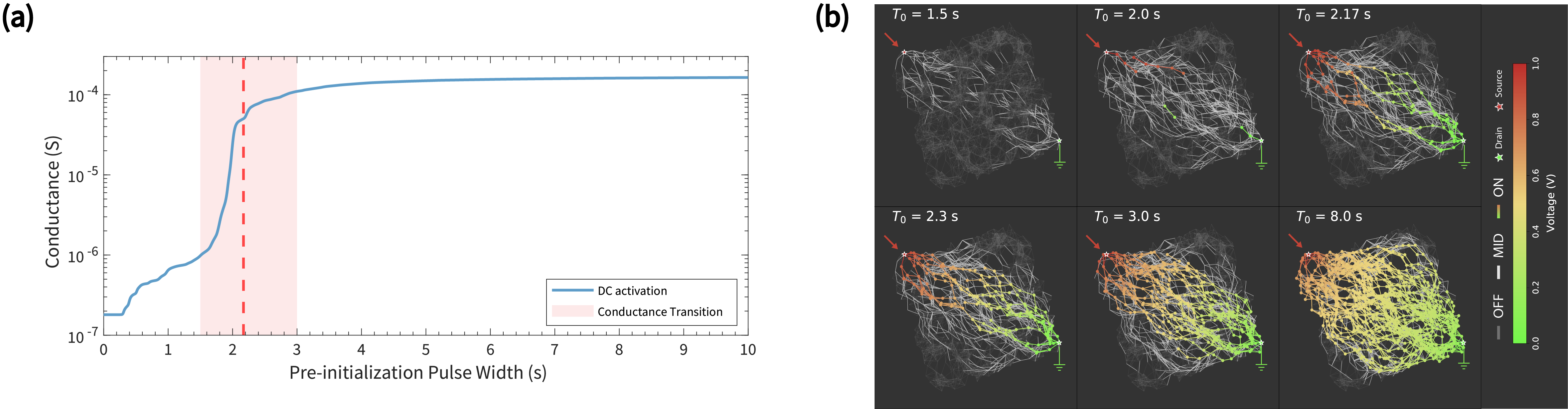}
    \caption{
    (a) Collective conductance of the network between the source and drain nodes with a DC input of varying pulse width. Dashed red line indicates the edge of formation of a high conductance current pathway between the source and drain nodes. 
    (b) Visualization of network activation levels for different pre-initilization pulse widths $T_0$. 
    }
    \label{fig:nw_pt}
\end{figure*}


For the purpose of using a NWN as a reservoir, a Mackey-Glass time-series signal with delay parameter $\tau = 17$ is delivered to a source electrode as the input voltage signal.
Before implementing the time-series prediction task using RC,
a DC input of varying duration is applied to the NWN to initialize the internal state of the network and prepare it for RC. We refer to this pre-initialization protocol as ``priming the system" \cite{zhurc}.
Fig.~\ref{fig:nw_pt}(a) shows the reservoir's conductance (blue curve) as a function of the DC input length $T_0$.
The shaded region represents the general conductance transition regime, identified from previous studies \cite{Hochstetter2021, Zhu2021information}, and the dashed line at $T_0=2.17\,$s 
represents when the first conductance pathways form between the source and drain nodes. 
The internal state of the network for different $T_0$ is visualized in Fig.~\ref{fig:nw_pt}(b).
In cases where the reservoir is under-activated ($T_0 < 2\,$s), the majority of memristive components remain inactive, resulting in insufficient dynamics from the network.
When the reservoir is over-activated ($T_0 > 8\,$s), the internal dynamics of the system become saturated, limiting the system's capacity to process additional information.
The conductance transition regime from Fig.~\ref{fig:nw_pt}(a) corresponds to an intermediate dynamical state of the reservoir, where conductance paths first span the network and the internal state of the system produces dynamical features that are more diverse than at other activation times.

\section{Ergodicity, reservoir computing and bistability}
\label{sec:RCandergodicity}

\subsection{Toy model}

Reservoir computing using~\cite{caravellisheldon} was studied for the first time in \cite{sheldonrc}, among other passive circuits. A brief recap of the procedures behind reservoir computing is provided in Supp. Mat. \ref{app:b}.

Here, we have used a similar scheme to the reservoir, using a memristive circuit comprising $N = 50$ idealized memristors.

We have considered parameter values $\alpha = \beta = 1$, and $\chi = 0.9$, for which the system experiences a symmetry break, and where we expect its dynamics to become non-ergodic, see~\ref{sec:memtoymodeltheory}. This is the regime where the reservoir is at the \textit{edge of stability},  as defined in~\cite{Carroll2020}, where in general, reservoirs can have optimal performance, although not always guaranteed. Weak ergodicity breaking in a dynamical system is associated with a strongly chaotic regime \cite{barkai}. We will come back to this point later.  We have also added a small noise value $\sigma = 0.01$. The equations~\eqref{eq:manydnoise} were numerically integrated with a time step $dt = 0.1$.

\begin{figure}[ht!]
    \centering
\includegraphics[width=.97\linewidth]{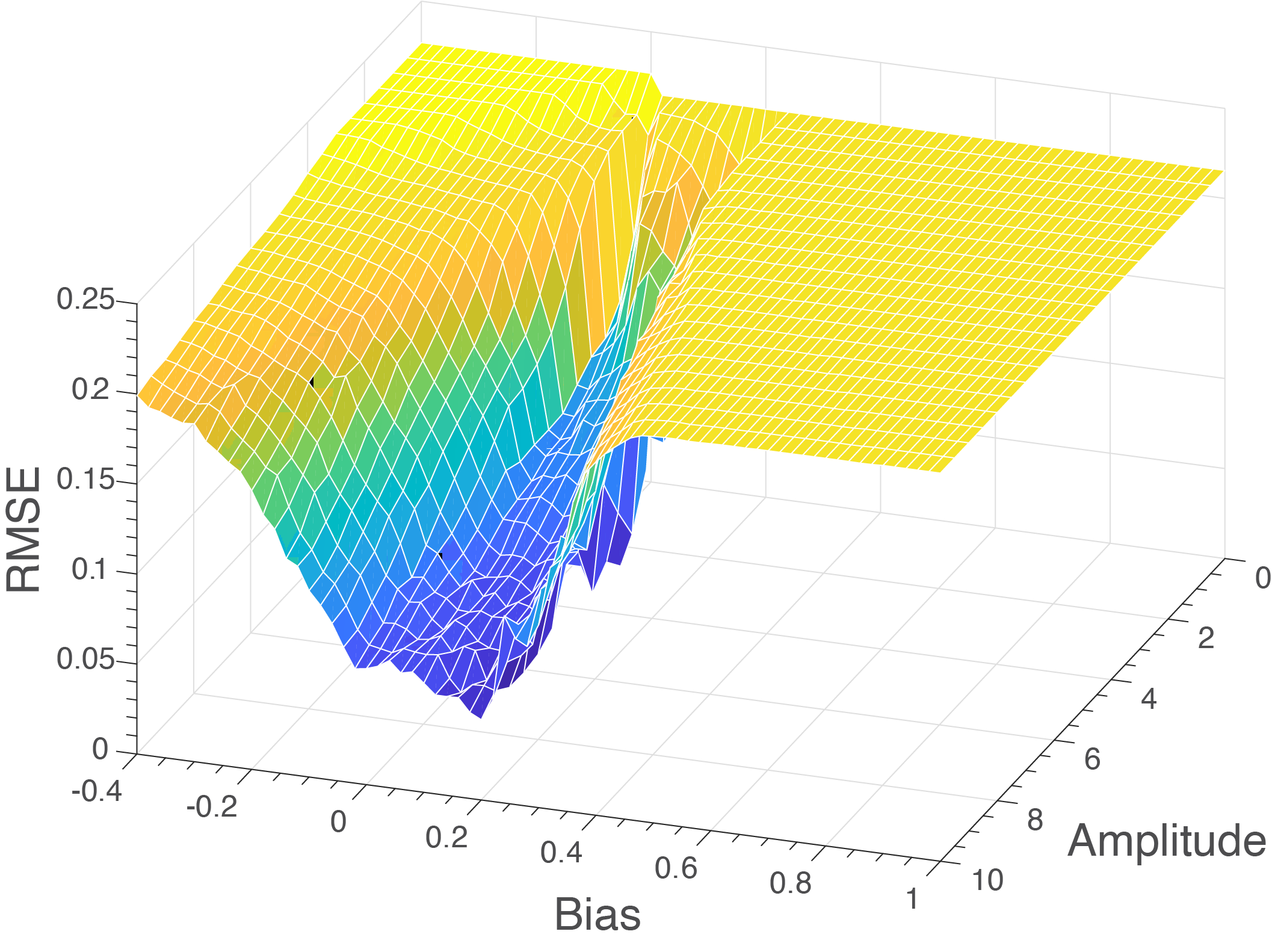}
     \caption{
     MSE of the Mackey-Glass reconstruction as a function of the parameters $a$ (Amp) and $b$ (bias).}
    \label{fig:MSE1tm}
\end{figure}

For the input signal $V(t)$ we have used a Mackey-Glass time series with delay parameter  $r = 0.2$, $\gamma = 0.1$, $\tau = 17$, and $n=10$.
Our input signal was given by
\begin{align}
    S(t) 
= 
    \mathrm{b}
+ 
    \mathrm{a} V(t),
\label{Eq:inputsignal}
\end{align}
with $ a \in [0,10]$ a multiplicative factor of the input $V(t)$ and $b \in [-0.4,1]$ a parameter bias. An example of the input signal is shown in Fig. \ref{fig:MSE1tm} (top). The parameter $b$ represents the bias input to the network, while $a$ is the amplitude.
%


For each value of $a$ and $b$, we performed a Mackey-Glass reconstruction task using the internal memory states as the dynamical variables for the RC. The rMSE, which we use as a measure of the performance of the RC, is shown in Fig. \ref{fig:MSE1tm}.
We can see from the figure that varying $a$ leads to a reduced value of the rMSE, right where the rumbling transition is present, approximately for $b = 0.2$.

\begin{figure}
    \centering
    \includegraphics[width=1\linewidth]{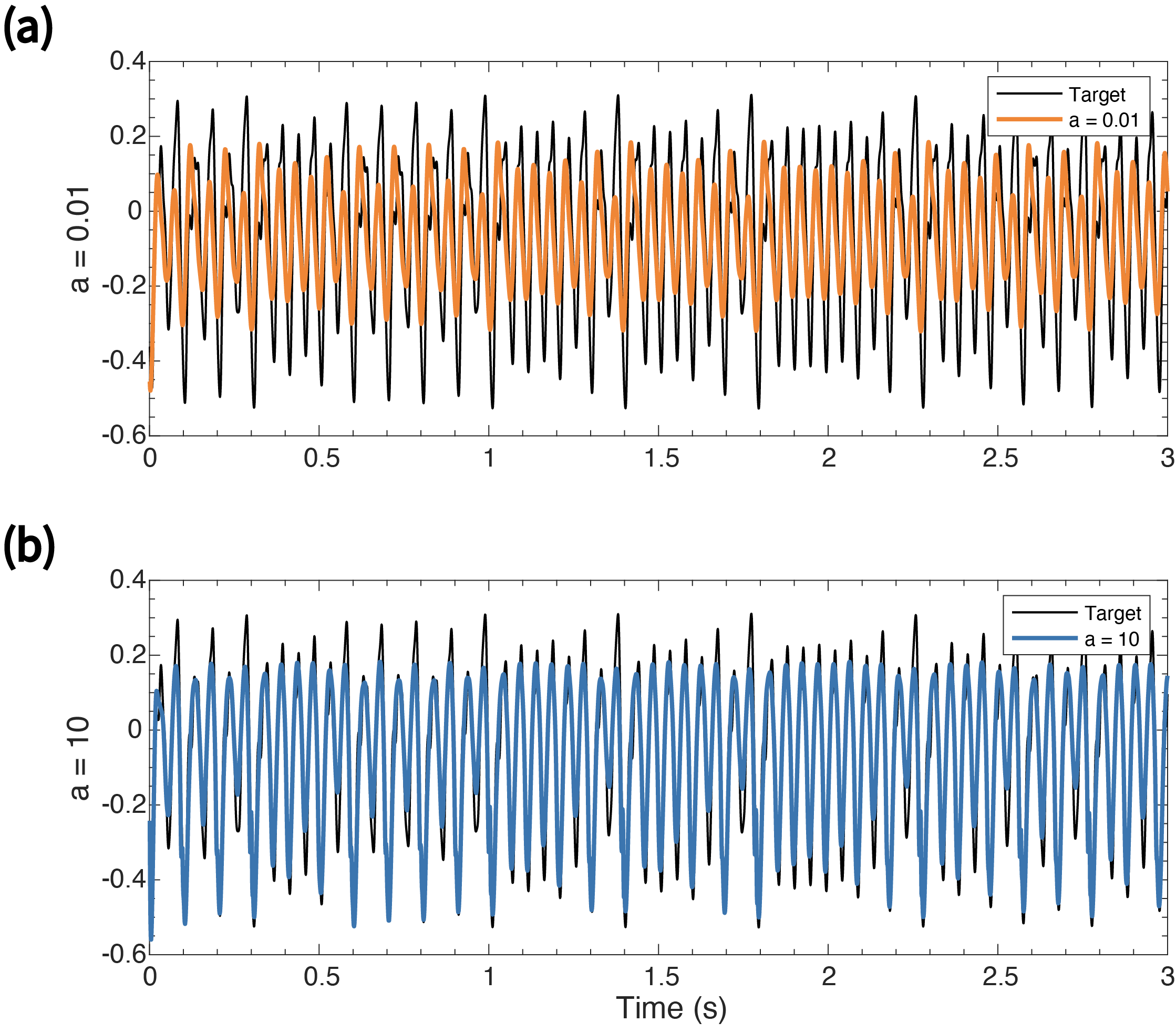}
    \caption{Reservoir computing prediction task with the toy model, for $b=0.2$, and $a=0.01$ (a) and $a=10$ (b)respectively. As we can see, the regime of optimal prediction corresponds both the regime in which the system's Thirumalai-Mountain metric does not converge to zero. }
    \label{fig:preddifftm}
\end{figure}

The difference in the quality of the task can be observed, for the Mackey-Glass time series, in Fig. \ref{fig:preddifftm}, for the toy model described in App. \ref{sec:tmod}.

\begin{figure}[ht!]
    \centering  
    \includegraphics[width=.97\linewidth]{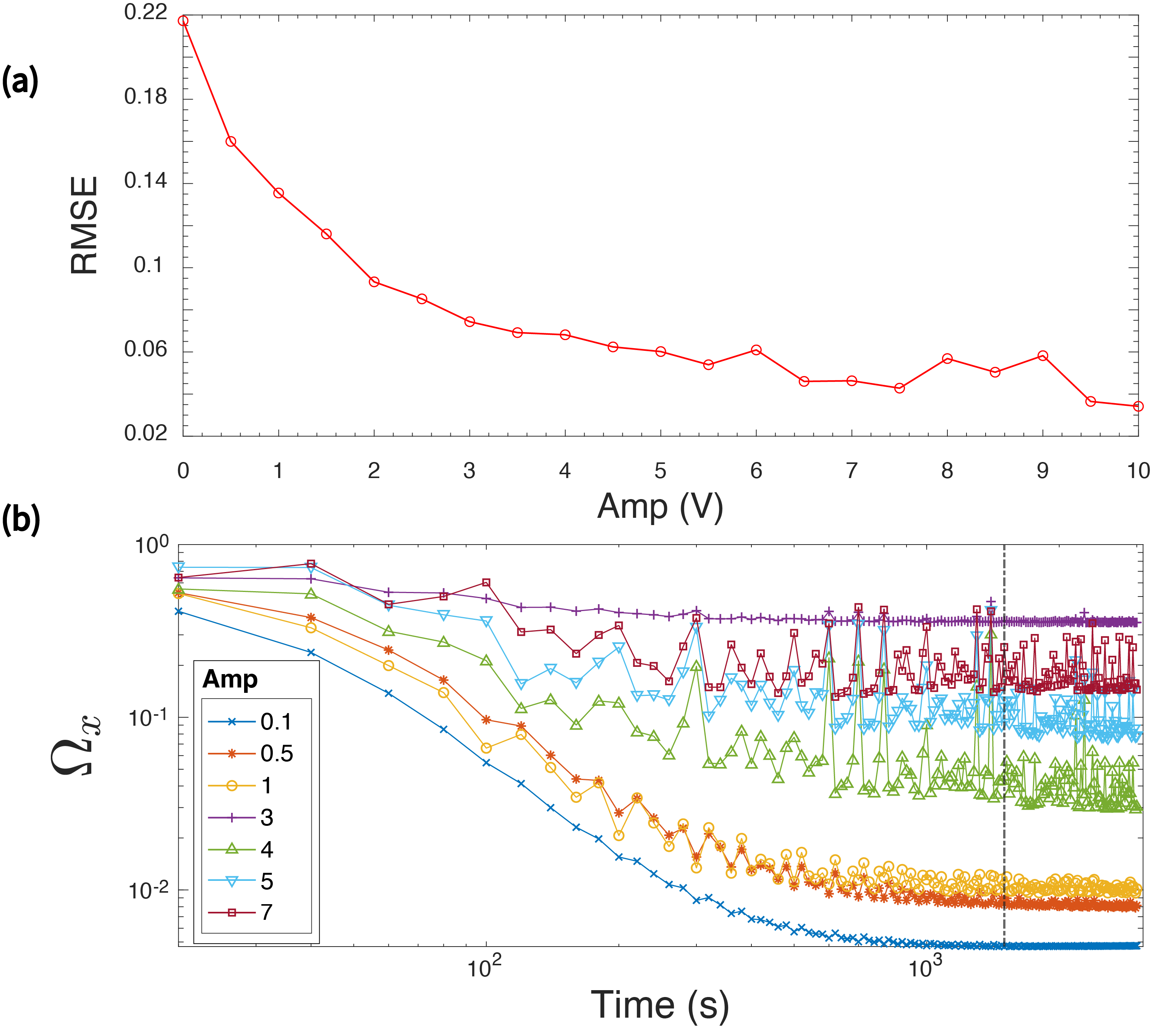}
    \caption{\textit{(a)}: Reservoir computing prediction error as a function of the amplitude of the input signal, while the bias is fixed on the transition point. We can see that rMSE decreases as a function of the amplitude. \textit{(a)}: Thirumalai-Mountain metric as a function of time for various amplitudes. We see that for increasing amplitudes, the metric ceases to converge to zero, indicating an effective non-ergodic behavior. In the intermediate regime, we see oscillations due to the fact that the system is jumping from one minimum to the other.
    }
    \label{fig:MSE-a02-b0-10}
\end{figure}

At a value of $b=0.2$, a plot of the rMSE is shown in Fig. \ref{fig:MSE-a02-b0-10}. As we can see, the value of the rMSE decreases as a function of $a$, meaning that the larger the magnitude of the input signal the better the performance, but this happens only when $b$ is carefully chosen near the transition point. The reason why this occurs can be inferred by analyzing the response of the internal memory values, which are reported in the Appendix, as a function of the parameter $b$ and for $a=0.2$, below, near, and above the transition point. When $a$ is such that the potential has a single minimum, the value of the memory oscillates in the vicinity of that minimum. When however $a$ is such that the potential has two minima, e.g. in the symmetry-breaking regime, at sufficiently large values of $b$ the memory values start to oscillate between the two minima. This implies that the time evolution of the system is much more dynamic in the symmetry-broken phase, and this effectively results in an improvement in the performance of the RC.

As we can see, increasing the value of the amplitude leads to the internal memory values fluctuating more prominently between the two stable states.
This is also shown in Fig. \ref{fig:MSE-a02-b0-10} (b), where we plot the TM metric evaluated on the response of the system in the small $Amp=1$ and large $Amp=10$ amplitude regimes. In one first case, the system is still ergodic, while in the case $Amp=10$ the TM metric is not converging to a value zero. 

Furthermore, we have calculated the $TM_x$ metric near the transition bias $a=0.2$ for varying values of the amplitude $b$ as 
\begin{align}
    \Omega_x(t)
=
    \frac{1}{N}
    \sum^N_j
    \left[
        \bar x_j(t)
    -
        \langle
            x
            (t)
        \rangle
    \right]^2
    .
\label{eq:TMtoymodelmeanfiel}
\end{align}
As the ensemble average  $\langle x(t) \rangle$ we have consider the mean field value $x_{cg}$ defined as~\cite{caravelli2021}
\begin{align}
    x_{cg}
=
    \frac{1}{N}
    \sum^N_{ij}
        {\mathcal P}_{ij}
        x_j
        .
\label{eq:meanfield}
\end{align}
We have chosen the mean field value $x_{cg}$ as a natural order parameter that resembles, in definition, the general meaning of ensemble average.

This is actually one of the advantages of studying this model first, as in this case, we know the details of the order parameter for the whole memristive network. 


\subsection{Nanowire networks}

The nanowire connectome dynamics has been extensively explored across various studies. Prior research has highlighted that nanowire Networks (NWNs) showcase brain-like dynamics, demonstrating their optimal information storage and processing capabilities at conductance transition points \cite{Hochstetter2021,Zhu2021information,caravelli2021}. More recently, a dynamical mean-field theoretical technique for polymer-coated Ag nanowires has uncovered emergent dynamical features \cite{Caravelli2023} such as transitions.  In the context of the two-terminal setup, these transitions are commonly observed within a regime termed as the 'edge of formation' \cite{Hochstetter2021}. This 'edge of formation' is delineated by the activation of memristive components, but precedes an exponential surge in the formation of parallel paths. As depicted in Fig.~\ref{fig:nw_pt}, this regime establishes a select few high-conductance current paths between the two electrodes (a phenomenon known as the 'winner-takes-all') \cite{wta}. Within this state, the internal dynamics of the NWN intricately map the input signal to a diverse feature space.

In the case of the toy model, previous analytical studies provide a comprehensive understanding of both the system dynamics and the order parameters to be employed \cite{Caravelli2016rl,caravelli2021}, which allows us to apply a simplification for the TM metric. 
Nevertheless, the same technique cannot be utilized for the nanowire model since it is more realistic and cannot be characterized in the same way.
For that reason, the collective conductance of the nanowire network is used as the order parameter, which is determined based on the individual conductances of individual memristive junctions and the underlying circuitry shaped by the connectivity.
In the meantime, the findings derived from the toy model will play an important role in interpreting the results of nanowires.

\subsubsection{Thirumalai-Mountain metric}
We now wish to link computation performance and ergodicity, also in the case of NWNs, we used the TM metric to understand the effective ergodicity of these dynamical systems;  in particular, eqn.~\ref{eq:TMtoymodelmeanfiel} is also used to calculate the TM metric for NWNs. However, the meaning associated with each quantity is closer to the definition of the metric used in the original work on supercooled liquid,~\cite{mountain89me}. The overall observable of interest is the two-probe conductance, defined earlier for nanowires, which is a scalar. Then, the time average is calculated using the global conductance, while the collective conductance is employed as an ensemble average across various realizations of the initial conditions:

\begin{align}
    \bar{g} (t) &= \frac{1}{T}\int_0^T G^i(t) dt ,\\
    \langle g(t) \rangle &= \frac{1}{N} \sum_{i}  G^i(t) ,
\end{align}
where $G^i(t)$ is the two-probe conductance of a particular realization at time $t$ and $G(t)$ is the collective and effective conductance of the network between the two points. For the time average quantity, note that we select a single element of the ensemble to evaluate the time average.

The ensemble of different realizations is generated by randomly perturbing the filament levels of junctions in the network:
\begin{align}
    \Lambda^i = (1 + \delta)\Lambda_0,
\end{align}
where $\delta$ is randomly sampled from a flat distribution, while $\Lambda^i$ is the parameter controlling the length of the filaments for all junctions, while $\Lambda_0$ is the initial condition of the filament. The random variable $\delta$ is sampled from a uniform distribution over $(-0.1, 0.1)$. 
Thus, effectively we are sampling over the initial conditions of the system.

\begin{figure}[h]
    \centering
    \includegraphics[width=.9\linewidth]{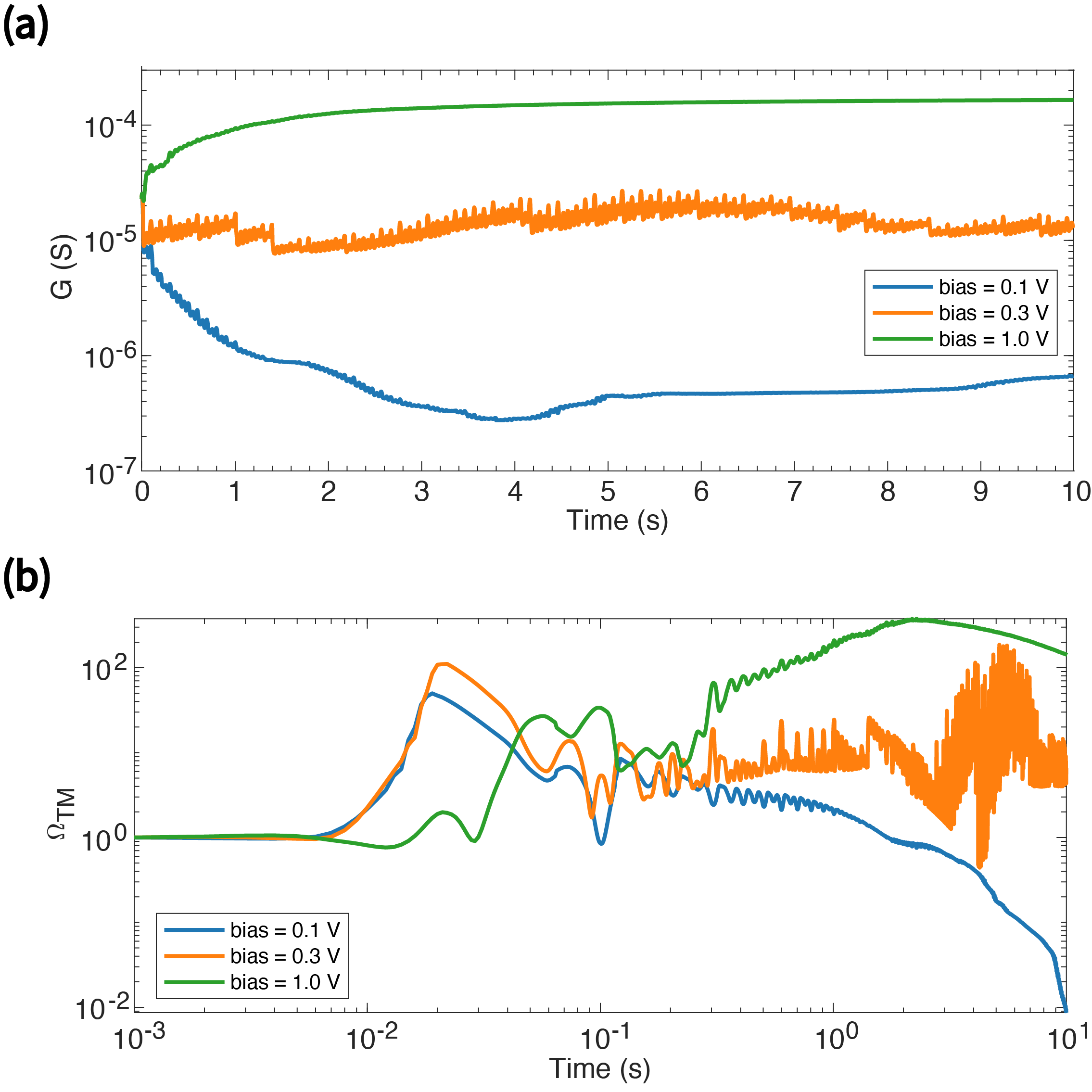}
    \caption{
    (a) Nanowire system simulated conductance, using the Mackey-Glass time series as input, for various values of the bias voltage $b$. As we can see, for small values and larger values of the bias, the conductance decays and grows to the asymptotic value. At intermediate values of the bias, the conductance oscillates.
    (b) Thirumalai-Mountain metric for the effective conductance of the nanowire network.}
    \label{fig:NW_G_TM}
\end{figure}

The TM metric, as a function of the bias, is shown in Fig. \ref{fig:NW_G_TM} (b). As we can see, and consistently with the case of the toy model, for small and large values of the bias, the metric decays. At intermediate values of the bias, at the edge of the transition between low and high conductance states, where the conductance synchronizes with the input, the TM metric fails to decay. This is exactly the regime in which the conductance transition occurs and is a signature of a special state for the nanowire network, highly synchronized with the input.



\subsubsection{Reservoir Computing}
For the realistic nanowire network model above, the implementation of RC involved the two probe conductance, with designated input and readout nodes.
Such construction has already been considered in the literature \cite{zhurc, Daniels2022}.

The fitting task of the Mackey-Glass time series was performed using NWNs under the RC framework, with two nanowire nodes in the network selected as the source and the drain. The MG signal was linearly transformed as described in Eq.~\ref{Eq:inputsignal} and delivered to the network as input, while the same MG signal with 5 steps ahead was employed as the target. The task can be breakdown into three phases:
\begin{enumerate}
    \item \textit{Priming:} A 2\,$V$ DC input of varying length $T_0$ was applied to drive the internal state of the network. The first 1000 data points of the MG time-series were delivered subsequently to wash out the influence from the initial state.

    \item \textit{Training:} The effective conductance of the network corresponding to $t = 1000 - 4000$\,ms was measured and multiplexed using the virtual node technique \cite{Appeltant2011} to provide training features (see details in Appendix). The readout layer was trained using a linear regression with a ridge parameter $r=0.01$.

    \item \textit{Testing:} The effective conductance during $t = 4000 - 5000$\,ms was measured and multiplexed in the same fashion as the training phase, while the trained readout layer was applied to make predictions. The performance of the reservoir can thus be evaluated accordingly.
\end{enumerate}



\begin{figure}[ht!]
    \centering
    \includegraphics[width=.99\linewidth]{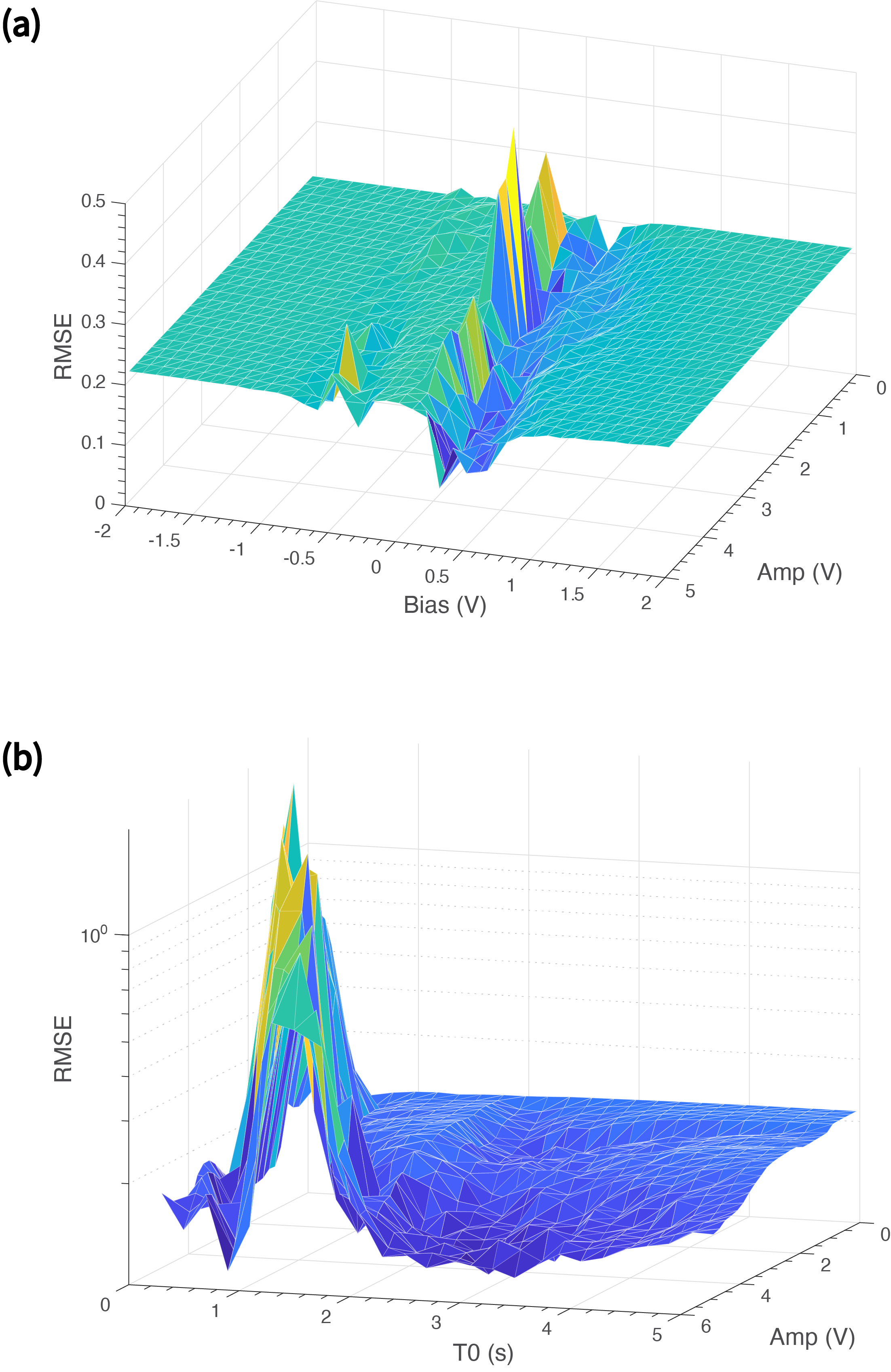}
    \caption{RMSE of RC prediction task with the simulated nanowire system as a function of driving. \textit{Top}: RMSE as a function of the bias and the amplitude of the input signal. As we can see, the minimum is located exactly at the numerically observed transition point in bias. \textit{Bottom}: $\log$RMSE as a function of the priming time $T0$ and the amplitude of the input. We see that the optimal results are obtained after a certain initial time $T0\approx 2$ s.  }
    \label{fig:priming}.
\end{figure}

\begin{figure}[h]
    \centering
    \includegraphics[width=.96\linewidth]{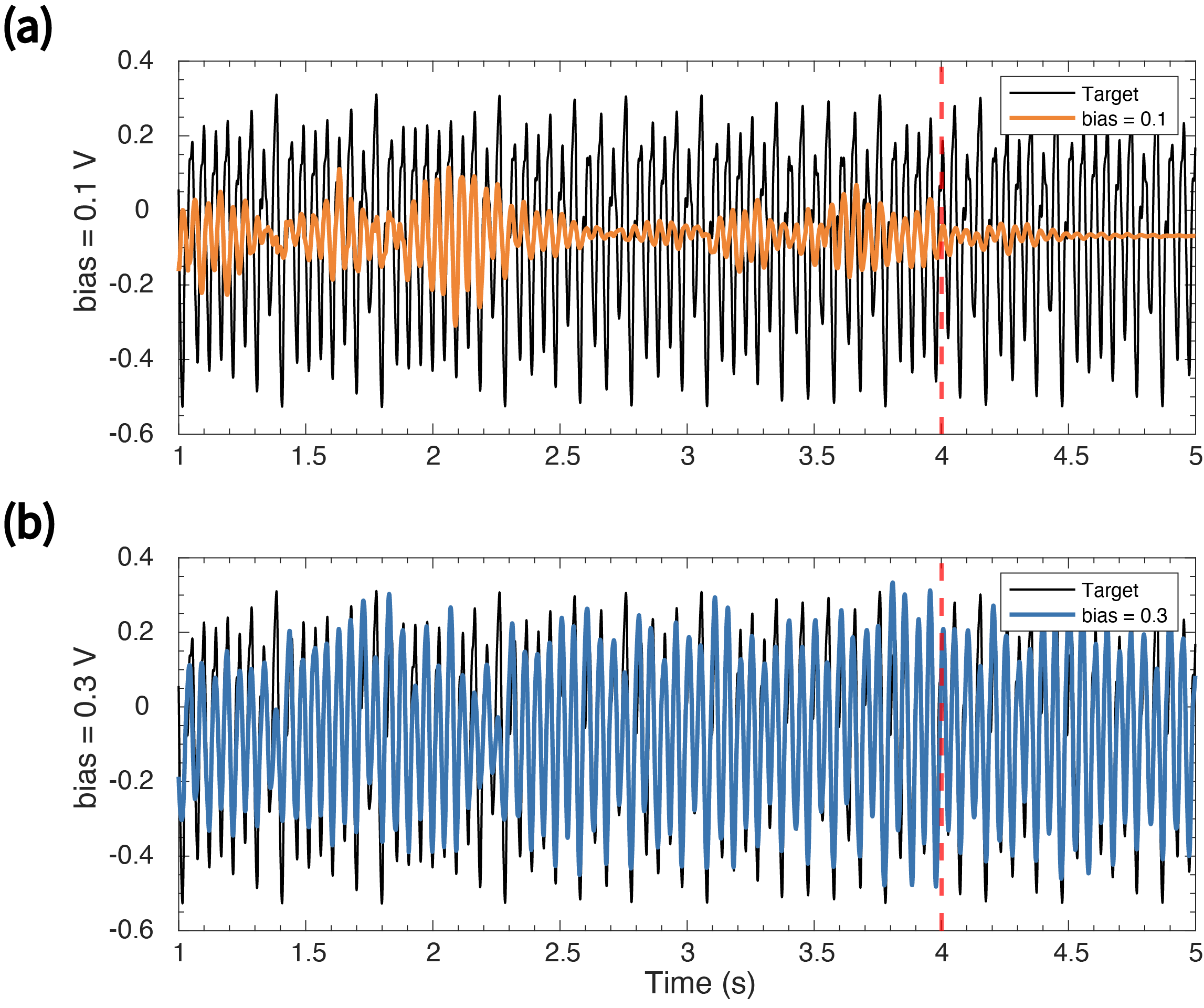}
    \caption{Fitting result for different biases, for amp = 4.9 V, $T_0$ = 2 s. As we can see, near the transition the performance of the RC prediction task increases dramatically. The red dashed line represents the divide between training and test sets.
    }
    \label{fig:fitfin}.
\end{figure}

We consider two regimes: one in which the system is initialized away from the voltage transition point between the high and low conductance state, and one in which the system sits at the boundary between the two.
As we can see from Fig. \ref{fig:priming} (top), the RMSE of the RC model has a behavior very similar to the one observed for the toy model of Fig. \ref{fig:MSE1tm}. At values of the amplitude close to the transition point, in which the Thirumalai-Mountain fails to converge to zero, the physical RC performance peaks. Our intuition is that the system tuned at the point in which it is at the edge of a transition, is more prone to synchronization \cite{Carroll2020} with the input signal. This, combined with existing knowledge \cite{zhurc} on the average ``priming" time that it takes for the system to reach synchronization with the signal, shows that the combination of input time and choice of input voltage leads to optimal performance. We can see the difference between the tuned and non-tuned network in Fig. \ref{fig:fitfin}.

\section{Conclusions}

Memory effects are a critical aspect of many physical and biological systems, and they have been shown to play a vital role in the behavior of complex systems. Meanwhile, ergodicity is a property of systems that describes the degree to which they explore their phase space. 
In recent years, in particular, physical systems with memory such as nanowires or nanoparticle connectomes, memristive, and other nanoscale devices, have become increasingly essential candidates as substrates for synthetic intelligent devices, e.g. brain-like physical material.
In particular, there are strong indications, both in theoretical models and experiments, that these devices exhibit conductance transitions both of the first and second order. These transitions have been linked to the edge of chaos, a concept that refers to the boundary between ordered and chaotic behavior in complex systems.

The present study explores the interplay between ergodicity breaking and memory in two models of memristive devices. The first model we studied was a toy model introduced in the literature to understand analytically the properties of purely memristive networks and has been instrumental in understanding their non-equilibrium properties such as Lyapunov functions \cite{Caravelli2019Ent} or conductance transitions \cite{caravelli2021}. The second is a more realistic model for memristive networks composed of nanowires \cite{Hochstetter2021,loeffler,Zhu2021information,zhurc}, used to understand the properties of polymer-coated Ag nanowires \cite{Diaz-Alvarez2019,milano001,Loeffler2023}. In both cases, we studied the Thirumalai-Mountain \cite{mountain89me} metric to understand how the systems relax when driven by different inputs, and in particular the ergodic properties of these systems. Both in the case of the toy model, for which the conductance transition has been studied analytically using a variety of techniques \cite{BCCV,BCCV2,caravelli2021,caravelliwein,caravelli2022}, and in the case of the nanowire model \cite{Hochstetter2021}, we found that the Thirumalai-Mountain metric signals lack of ergodicity near the voltage bias value where the conductance transitions are expected to be.

This is, in particular, interesting in view of the fact that it has been recently observed that edge of instability can be linked to a computational advantage \cite{edgeofchaos,Carroll2020}. A similar result is observed in this paper. We did observe that, in particular, this is not necessarily true unless the dynamical system under study is synchronized to the input signal, as suggested in \cite{Carroll2020}. In fact, this improved performance occurs only after a transient period (``priming") which allows the nanowire network to synchronize to the input signal, as shown in Fig. \ref{fig:priming}. The results described above are consistent across two different models: the toy model in which conductance transitions can be understood quantitatively and analytically, and the more realistic nanowire model able to capture the experimentally observed conductance in Ag nanowires. Thus, these results suggest that there might be an underlying common theory to explain these transitions.

In conclusion, by connecting memory effects, ergodicity, and the edge of chaos, we have identified a set of principles that can be used to create more effective computational models. Our research has shown for the first time that non-ergodic behavior can be linked to the effectiveness of reservoir computing, leading to new approaches for developing more advanced and efficient computational tools. We believe that our findings will inspire further investigations into the connections between memory effects, ergodicity, and the edge of chaos, and will lead to new and exciting developments in the field of machine learning and beyond. In particular, in future work we will discuss the application of the ideas developed in this paper to meta-plasticity with memristive systems, \cite{Chialvo1999,Carbajal2022}, in particular in view of the recent results on the meta-plasticity of nanowire networks shown in \cite{Loeffler2023}.

\begin{acknowledgements}
The work of FC was carried out under the auspices of the NNSA of the U.S. DoE at LANL under Contract No. DE-AC52-06NA25396, and in particular support from LDRD via 20230338ER and 20230627ER.
RZ is supported by a Postgraduate Research Excellence Award scholarship from the University of Sydney. VB acknowledges funding through the RMIT Vice-Chancelor’s Research Fellowship.
\end{acknowledgements}
\bibliography{bibliography}

\begin{thebibliography}{77}%
\makeatletter
\providecommand \@ifxundefined [1]{%
 \@ifx{#1\undefined}
}%
\providecommand \@ifnum [1]{%
 \ifnum #1\expandafter \@firstoftwo
 \else \expandafter \@secondoftwo
 \fi
}%
\providecommand \@ifx [1]{%
 \ifx #1\expandafter \@firstoftwo
 \else \expandafter \@secondoftwo
 \fi
}%
\providecommand \natexlab [1]{#1}%
\providecommand \enquote  [1]{``#1''}%
\providecommand \bibnamefont  [1]{#1}%
\providecommand \bibfnamefont [1]{#1}%
\providecommand \citenamefont [1]{#1}%
\providecommand \href@noop [0]{\@secondoftwo}%
\providecommand \href [0]{\begingroup \@sanitize@url \@href}%
\providecommand \@href[1]{\@@startlink{#1}\@@href}%
\providecommand \@@href[1]{\endgroup#1\@@endlink}%
\providecommand \@sanitize@url [0]{\catcode `\\12\catcode `\$12\catcode
  `\&12\catcode `\#12\catcode `\^12\catcode `\_12\catcode `\%12\relax}%
\providecommand \@@startlink[1]{}%
\providecommand \@@endlink[0]{}%
\providecommand \url  [0]{\begingroup\@sanitize@url \@url }%
\providecommand \@url [1]{\endgroup\@href {#1}{\urlprefix }}%
\providecommand \urlprefix  [0]{URL }%
\providecommand \Eprint [0]{\href }%
\providecommand \doibase [0]{https://doi.org/}%
\providecommand \selectlanguage [0]{\@gobble}%
\providecommand \bibinfo  [0]{\@secondoftwo}%
\providecommand \bibfield  [0]{\@secondoftwo}%
\providecommand \translation [1]{[#1]}%
\providecommand \BibitemOpen [0]{}%
\providecommand \bibitemStop [0]{}%
\providecommand \bibitemNoStop [0]{.\EOS\space}%
\providecommand \EOS [0]{\spacefactor3000\relax}%
\providecommand \BibitemShut  [1]{\csname bibitem#1\endcsname}%
\let\auto@bib@innerbib\@empty
\bibitem [{\citenamefont {Mehonic}\ \emph {et~al.}(2020)\citenamefont
  {Mehonic}, \citenamefont {Sebastian}, \citenamefont {Rajendran},
  \citenamefont {Simeone}, \citenamefont {Vasilaki},\ and\ \citenamefont
  {Kenyon}}]{Mehonic2020}%
  \BibitemOpen
  \bibfield  {author} {\bibinfo {author} {\bibfnamefont {A.}~\bibnamefont
  {Mehonic}}, \bibinfo {author} {\bibfnamefont {A.}~\bibnamefont {Sebastian}},
  \bibinfo {author} {\bibfnamefont {B.}~\bibnamefont {Rajendran}}, \bibinfo
  {author} {\bibfnamefont {O.}~\bibnamefont {Simeone}}, \bibinfo {author}
  {\bibfnamefont {E.}~\bibnamefont {Vasilaki}},\ and\ \bibinfo {author}
  {\bibfnamefont {A.~J.}\ \bibnamefont {Kenyon}},\ }\bibfield  {title}
  {\bibinfo {title} {Memristors—from in-memory computing, deep learning
  acceleration, and spiking neural networks to the future of neuromorphic and
  bio-inspired computing},\ }\href
  {https://doi.org/https://doi.org/10.1002/aisy.202000085} {\bibfield
  {journal} {\bibinfo  {journal} {Advanced Intelligent Systems}\ }\textbf
  {\bibinfo {volume} {2}},\ \bibinfo {pages} {2000085} (\bibinfo {year}
  {2020})}\BibitemShut {NoStop}%
\bibitem [{\citenamefont {Oliver}(2019)}]{Oliver2019}%
  \BibitemOpen
  \bibfield  {author} {\bibinfo {author} {\bibfnamefont {W.~D.}\ \bibnamefont
  {Oliver}},\ }\bibfield  {title} {\bibinfo {title} {Quantum computing takes
  flight},\ }\href {https://doi.org/10.1038/d41586-019-03173-4} {\bibfield
  {journal} {\bibinfo  {journal} {Nature}\ }\textbf {\bibinfo {volume} {574}},\
  \bibinfo {pages} {487} (\bibinfo {year} {2019})}\BibitemShut {NoStop}%
\bibitem [{\citenamefont {B\"{o}hm}\ \emph {et~al.}(2019)\citenamefont
  {B\"{o}hm}, \citenamefont {Verschaffelt},\ and\ \citenamefont {der
  Sande}}]{Isingmachine}%
  \BibitemOpen
  \bibfield  {author} {\bibinfo {author} {\bibfnamefont {F.}~\bibnamefont
  {B\"{o}hm}}, \bibinfo {author} {\bibfnamefont {G.}~\bibnamefont
  {Verschaffelt}},\ and\ \bibinfo {author} {\bibfnamefont {G.~V.}\ \bibnamefont
  {der Sande}},\ }\bibfield  {title} {\bibinfo {title} {A poor man's coherent
  ising machine based on opto-electronic feedback systems for solving
  optimization problems},\ }\bibfield  {journal} {\bibinfo  {journal} {Nature
  Comm.}\ }\textbf {\bibinfo {volume} {10}},\ \href
  {https://doi.org/10.1038/s41467-019-11484-3} {10.1038/s41467-019-11484-3}
  (\bibinfo {year} {2019})\BibitemShut {NoStop}%
\bibitem [{\citenamefont {Pierangeli}\ \emph {et~al.}(2019)\citenamefont
  {Pierangeli}, \citenamefont {Marcucci},\ and\ \citenamefont
  {Conti}}]{Pierangeli_2019}%
  \BibitemOpen
  \bibfield  {author} {\bibinfo {author} {\bibfnamefont {D.}~\bibnamefont
  {Pierangeli}}, \bibinfo {author} {\bibfnamefont {G.}~\bibnamefont
  {Marcucci}},\ and\ \bibinfo {author} {\bibfnamefont {C.}~\bibnamefont
  {Conti}},\ }\bibfield  {title} {\bibinfo {title} {Large-scale photonic ising
  machine by spatial light modulation},\ }\href
  {https://doi.org/10.1103/physrevlett.122.213902} {\bibfield  {journal}
  {\bibinfo  {journal} {Phys. Rev. Lett.}\ }\textbf {\bibinfo {volume} {122}},\
  \bibinfo {pages} {213902} (\bibinfo {year} {2019})}\BibitemShut {NoStop}%
\bibitem [{\citenamefont {Vadlamani}\ \emph {et~al.}(2020)\citenamefont
  {Vadlamani}, \citenamefont {Xiao},\ and\ \citenamefont
  {Yablonovitch}}]{Vadlamani2020}%
  \BibitemOpen
  \bibfield  {author} {\bibinfo {author} {\bibfnamefont {S.~K.}\ \bibnamefont
  {Vadlamani}}, \bibinfo {author} {\bibfnamefont {T.~P.}\ \bibnamefont
  {Xiao}},\ and\ \bibinfo {author} {\bibfnamefont {E.}~\bibnamefont
  {Yablonovitch}},\ }\bibfield  {title} {\bibinfo {title} {Physics successfully
  implements lagrange multiplier optimization},\ }\href
  {https://doi.org/10.1073/pnas.2015192117} {\bibfield  {journal} {\bibinfo
  {journal} {Proc. of the Nat. Aca. of Sci.}\ }\textbf {\bibinfo {volume}
  {117}},\ \bibinfo {pages} {26639} (\bibinfo {year} {2020})}\BibitemShut
  {NoStop}%
\bibitem [{\citenamefont {Csaba}\ and\ \citenamefont
  {Porod}(2020)}]{Csaba2020}%
  \BibitemOpen
  \bibfield  {author} {\bibinfo {author} {\bibfnamefont {G.}~\bibnamefont
  {Csaba}}\ and\ \bibinfo {author} {\bibfnamefont {W.}~\bibnamefont {Porod}},\
  }\bibfield  {title} {\bibinfo {title} {Coupled oscillators for computing: A
  review and perspective},\ }\href {https://doi.org/10.1063/1.5120412}
  {\bibfield  {journal} {\bibinfo  {journal} {App. Phys. Rev.}\ }\textbf
  {\bibinfo {volume} {7}},\ \bibinfo {pages} {011302} (\bibinfo {year}
  {2020})}\BibitemShut {NoStop}%
\bibitem [{\citenamefont {Goto}\ and\ \citenamefont {et~al.}(2021)}]{goto}%
  \BibitemOpen
  \bibfield  {author} {\bibinfo {author} {\bibfnamefont {H.}~\bibnamefont
  {Goto}}\ and\ \bibinfo {author} {\bibnamefont {et~al.}},\ }\bibfield  {title}
  {\bibinfo {title} {High-performance combinatorial optimization based on
  classical mechanics},\ }\href@noop {} {\bibfield  {journal} {\bibinfo
  {journal} {Sci. Adv.}\ } (\bibinfo {year} {2021})}\BibitemShut {NoStop}%
\bibitem [{\citenamefont {Singh}\ and\ \citenamefont {et.
  al.}(2019)}]{Singh2019}%
  \BibitemOpen
  \bibfield  {author} {\bibinfo {author} {\bibfnamefont {G.}~\bibnamefont
  {Singh}}\ and\ \bibinfo {author} {\bibnamefont {et. al.}},\ }\bibfield
  {title} {\bibinfo {title} {Near-memory computing: Past, present, and
  future},\ }\href {https://doi.org/10.1016/j.micpro.2019.102868} {\bibfield
  {journal} {\bibinfo  {journal} {Micro. and Micro.}\ }\textbf {\bibinfo
  {volume} {71}},\ \bibinfo {pages} {102868} (\bibinfo {year}
  {2019})}\BibitemShut {NoStop}%
\bibitem [{\citenamefont {Ielmini}\ and\ \citenamefont
  {Wong}(2018)}]{Ielmini2018}%
  \BibitemOpen
  \bibfield  {author} {\bibinfo {author} {\bibfnamefont {D.}~\bibnamefont
  {Ielmini}}\ and\ \bibinfo {author} {\bibfnamefont {H.-S.~P.}\ \bibnamefont
  {Wong}},\ }\bibfield  {title} {\bibinfo {title} {In-memory computing with
  resistive switching devices},\ }\href
  {https://doi.org/10.1038/s41928-018-0092-2} {\bibfield  {journal} {\bibinfo
  {journal} {Nature Ele.}\ }\textbf {\bibinfo {volume} {1}},\ \bibinfo {pages}
  {333} (\bibinfo {year} {2018})}\BibitemShut {NoStop}%
\bibitem [{\citenamefont {Traversa}\ and\ \citenamefont {et.
  al.}(2014)}]{DCRAM}%
  \BibitemOpen
  \bibfield  {author} {\bibinfo {author} {\bibfnamefont {F.~L.}\ \bibnamefont
  {Traversa}}\ and\ \bibinfo {author} {\bibnamefont {et. al.}},\ }\bibfield
  {title} {\bibinfo {title} {Dynamic computing random access memory},\
  }\href@noop {} {\bibfield  {journal} {\bibinfo  {journal} {Nanotechnology}\
  }\textbf {\bibinfo {volume} {25}},\ \bibinfo {pages} {285201} (\bibinfo
  {year} {2014})}\BibitemShut {NoStop}%
\bibitem [{\citenamefont {Sebastian}\ and\ \citenamefont {et.
  al.}(2020)}]{Sebastian2020}%
  \BibitemOpen
  \bibfield  {author} {\bibinfo {author} {\bibfnamefont {A.}~\bibnamefont
  {Sebastian}}\ and\ \bibinfo {author} {\bibnamefont {et. al.}},\ }\bibfield
  {title} {\bibinfo {title} {Memory devices and applications for in-memory
  computing},\ }\href {https://doi.org/10.1038/s41565-020-0655-z} {\bibfield
  {journal} {\bibinfo  {journal} {Nature Nano.}\ }\textbf {\bibinfo {volume}
  {15}},\ \bibinfo {pages} {529} (\bibinfo {year} {2020})}\BibitemShut
  {NoStop}%
\bibitem [{\citenamefont {Traversa}\ and\ \citenamefont {{Di
  Ventra}}(2015)}]{traversa}%
  \BibitemOpen
  \bibfield  {author} {\bibinfo {author} {\bibfnamefont {F.~L.}\ \bibnamefont
  {Traversa}}\ and\ \bibinfo {author} {\bibfnamefont {M.}~\bibnamefont {{Di
  Ventra}}},\ }\bibfield  {title} {\bibinfo {title} {Universal memcomputing
  machines},\ }\href {https://doi.org/10.1109/tnnls.2015.2391182} {\bibfield
  {journal} {\bibinfo  {journal} {{IEEE} Trans. on Neur. Net. and Learn. Sys.}\
  }\textbf {\bibinfo {volume} {26}},\ \bibinfo {pages} {2702} (\bibinfo {year}
  {2015})}\BibitemShut {NoStop}%
\bibitem [{\citenamefont {{Di Ventra}}\ and\ \citenamefont
  {Traversa}(2018)}]{Ventra2018}%
  \BibitemOpen
  \bibfield  {author} {\bibinfo {author} {\bibfnamefont {M.}~\bibnamefont {{Di
  Ventra}}}\ and\ \bibinfo {author} {\bibfnamefont {F.~L.}\ \bibnamefont
  {Traversa}},\ }\bibfield  {title} {\bibinfo {title} {Perspective:
  Memcomputing: Leveraging memory and physics to compute efficiently},\ }\href
  {https://doi.org/10.1063/1.5026506} {\bibfield  {journal} {\bibinfo
  {journal} {J. of App. Phys.}\ }\textbf {\bibinfo {volume} {123}},\ \bibinfo
  {pages} {180901} (\bibinfo {year} {2018})}\BibitemShut {NoStop}%
\bibitem [{\citenamefont {Hennessy}\ and\ \citenamefont
  {Patterson}(2019)}]{Hennessy2019}%
  \BibitemOpen
  \bibfield  {author} {\bibinfo {author} {\bibfnamefont {J.~L.}\ \bibnamefont
  {Hennessy}}\ and\ \bibinfo {author} {\bibfnamefont {D.~A.}\ \bibnamefont
  {Patterson}},\ }\bibfield  {title} {\bibinfo {title} {A new golden age for
  computer architecture},\ }\href {https://doi.org/10.1145/3282307} {\bibfield
  {journal} {\bibinfo  {journal} {Comm. of the {ACM}}\ }\textbf {\bibinfo
  {volume} {62}},\ \bibinfo {pages} {48} (\bibinfo {year} {2019})}\BibitemShut
  {NoStop}%
\bibitem [{\citenamefont {Sutton}\ and\ \citenamefont {et.
  al.}(2017)}]{Sutton2017}%
  \BibitemOpen
  \bibfield  {author} {\bibinfo {author} {\bibfnamefont {B.}~\bibnamefont
  {Sutton}}\ and\ \bibinfo {author} {\bibnamefont {et. al.}},\ }\bibfield
  {title} {\bibinfo {title} {Intrinsic optimization using stochastic
  nanomagnets},\ }\bibfield  {journal} {\bibinfo  {journal} {Sci. Rep.}\
  }\textbf {\bibinfo {volume} {7}},\ \href {https://doi.org/10.1038/srep44370}
  {10.1038/srep44370} (\bibinfo {year} {2017})\BibitemShut {NoStop}%
\bibitem [{\citenamefont {Torrejon}\ and\ \citenamefont {et.
  al.}(2017)}]{Torrejon2017}%
  \BibitemOpen
  \bibfield  {author} {\bibinfo {author} {\bibfnamefont {J.}~\bibnamefont
  {Torrejon}}\ and\ \bibinfo {author} {\bibnamefont {et. al.}},\ }\bibfield
  {title} {\bibinfo {title} {Neuromorphic computing with nanoscale spintronic
  oscillators},\ }\href {https://doi.org/10.1038/nature23011} {\bibfield
  {journal} {\bibinfo  {journal} {Nature}\ }\textbf {\bibinfo {volume} {547}},\
  \bibinfo {pages} {428} (\bibinfo {year} {2017})}\BibitemShut {NoStop}%
\bibitem [{\citenamefont {Kirkpatrick}\ \emph {et~al.}(1983)\citenamefont
  {Kirkpatrick}, \citenamefont {Gelatt},\ and\ \citenamefont
  {Vecchi}}]{Kirkpatrick83}%
  \BibitemOpen
  \bibfield  {author} {\bibinfo {author} {\bibfnamefont {S.}~\bibnamefont
  {Kirkpatrick}}, \bibinfo {author} {\bibfnamefont {C.~D.}\ \bibnamefont
  {Gelatt}},\ and\ \bibinfo {author} {\bibfnamefont {M.~P.}\ \bibnamefont
  {Vecchi}},\ }\bibfield  {title} {\bibinfo {title} {Optimization by simulated
  annealing},\ }\href {https://doi.org/10.1126/science.220.4598.671} {\bibfield
   {journal} {\bibinfo  {journal} {Science}\ }\textbf {\bibinfo {volume}
  {220}},\ \bibinfo {pages} {671} (\bibinfo {year} {1983})}\BibitemShut
  {NoStop}%
\bibitem [{\citenamefont {Dorigo}\ and\ \citenamefont
  {Stützle}(2004)}]{Dorigo2004}%
  \BibitemOpen
  \bibfield  {author} {\bibinfo {author} {\bibfnamefont {M.}~\bibnamefont
  {Dorigo}}\ and\ \bibinfo {author} {\bibfnamefont {T.}~\bibnamefont
  {Stützle}},\ }\href {https://doi.org/10.7551/mitpress/1290.001.0001} {\emph
  {\bibinfo {title} {Ant Colony Optimization}}}\ (\bibinfo  {publisher}
  {{MIT}},\ \bibinfo {year} {2004})\BibitemShut {NoStop}%
\bibitem [{\citenamefont {Milano}\ \emph {et~al.}(2019)\citenamefont {Milano},
  \citenamefont {Porro}, \citenamefont {Valov},\ and\ \citenamefont
  {Ricciardi}}]{Milano2019}%
  \BibitemOpen
  \bibfield  {author} {\bibinfo {author} {\bibfnamefont {G.}~\bibnamefont
  {Milano}}, \bibinfo {author} {\bibfnamefont {S.}~\bibnamefont {Porro}},
  \bibinfo {author} {\bibfnamefont {I.}~\bibnamefont {Valov}},\ and\ \bibinfo
  {author} {\bibfnamefont {C.}~\bibnamefont {Ricciardi}},\ }\bibfield  {title}
  {\bibinfo {title} {Recent developments and perspectives for memristive
  devices based on metal oxide nanowires},\ }\href
  {https://doi.org/https://doi.org/10.1002/aelm.201800909} {\bibfield
  {journal} {\bibinfo  {journal} {Advanced Electronic Materials}\ }\textbf
  {\bibinfo {volume} {5}},\ \bibinfo {pages} {1800909} (\bibinfo {year}
  {2019})}\BibitemShut {NoStop}%
\bibitem [{\citenamefont {Kuncic}\ and\ \citenamefont
  {Nakayama}(2021)}]{Kuncic2021}%
  \BibitemOpen
  \bibfield  {author} {\bibinfo {author} {\bibfnamefont {Z.}~\bibnamefont
  {Kuncic}}\ and\ \bibinfo {author} {\bibfnamefont {T.}~\bibnamefont
  {Nakayama}},\ }\bibfield  {title} {\bibinfo {title} {Neuromorphic nanowire
  networks: principles, progress and future prospects for neuro-inspired
  information processing},\ }\href
  {https://doi.org/10.1080/23746149.2021.1894234} {\bibfield  {journal}
  {\bibinfo  {journal} {Advances in Physics: X}\ }\textbf {\bibinfo {volume}
  {6}},\ \bibinfo {pages} {1894234} (\bibinfo {year} {2021})}\BibitemShut
  {NoStop}%
\bibitem [{\citenamefont {Hochstetter}\ and\ \citenamefont
  {et~al.}(2021)}]{Hochstetter2021}%
  \BibitemOpen
  \bibfield  {author} {\bibinfo {author} {\bibfnamefont {J.}~\bibnamefont
  {Hochstetter}}\ and\ \bibinfo {author} {\bibnamefont {et~al.}},\ }\bibfield
  {title} {\bibinfo {title} {Avalanches and edge-of-chaos learning in
  neuromorphic nanowire networks},\ }\bibfield  {journal} {\bibinfo  {journal}
  {Nature Comm.}\ }\textbf {\bibinfo {volume} {12}},\ \href
  {https://doi.org/10.1038/s41467-021-24260-z} {10.1038/s41467-021-24260-z}
  (\bibinfo {year} {2021})\BibitemShut {NoStop}%
\bibitem [{\citenamefont {{Diaz-Alvarez}}\ \emph {et~al.}(2019)\citenamefont
  {{Diaz-Alvarez}}, \citenamefont {Higuchi}, \citenamefont {{Sanz-Leon}},
  \citenamefont {Marcus}, \citenamefont {Shingaya}, \citenamefont {Stieg},
  \citenamefont {Gimzewski}, \citenamefont {Kuncic},\ and\ \citenamefont
  {Nakayama}}]{Diaz-Alvarez2019}%
  \BibitemOpen
  \bibfield  {author} {\bibinfo {author} {\bibfnamefont {A.}~\bibnamefont
  {{Diaz-Alvarez}}}, \bibinfo {author} {\bibfnamefont {R.}~\bibnamefont
  {Higuchi}}, \bibinfo {author} {\bibfnamefont {P.}~\bibnamefont
  {{Sanz-Leon}}}, \bibinfo {author} {\bibfnamefont {I.}~\bibnamefont {Marcus}},
  \bibinfo {author} {\bibfnamefont {Y.}~\bibnamefont {Shingaya}}, \bibinfo
  {author} {\bibfnamefont {A.~Z.}\ \bibnamefont {Stieg}}, \bibinfo {author}
  {\bibfnamefont {J.~K.}\ \bibnamefont {Gimzewski}}, \bibinfo {author}
  {\bibfnamefont {Z.}~\bibnamefont {Kuncic}},\ and\ \bibinfo {author}
  {\bibfnamefont {T.}~\bibnamefont {Nakayama}},\ }\bibfield  {title} {\bibinfo
  {title} {Emergent dynamics of neuromorphic nanowire networks},\ }\href
  {https://doi.org/10.1038/s41598-019-51330-6} {\bibfield  {journal} {\bibinfo
  {journal} {Scientific Reports}\ }\textbf {\bibinfo {volume} {9}},\ \bibinfo
  {pages} {14920} (\bibinfo {year} {2019})}\BibitemShut {NoStop}%
\bibitem [{\citenamefont {Caravelli}\ \emph {et~al.}(2021)\citenamefont
  {Caravelli}, \citenamefont {Sheldon},\ and\ \citenamefont
  {Traversa}}]{caravelli2021}%
  \BibitemOpen
  \bibfield  {author} {\bibinfo {author} {\bibfnamefont {F.}~\bibnamefont
  {Caravelli}}, \bibinfo {author} {\bibfnamefont {F.}~\bibnamefont {Sheldon}},\
  and\ \bibinfo {author} {\bibfnamefont {F.~L.}\ \bibnamefont {Traversa}},\
  }\bibfield  {title} {\bibinfo {title} {Global minimization via classical
  tunneling assisted by collective force field formation},\ }\href
  {https://doi.org/10.1126/sciadv.abh1542} {\bibfield  {journal} {\bibinfo
  {journal} {Science Advances}\ }\textbf {\bibinfo {volume} {7}},\ \bibinfo
  {pages} {1542} (\bibinfo {year} {2021})},\ \Eprint
  {https://arxiv.org/abs/https://www.science.org/doi/pdf/10.1126/sciadv.abh1542}
  {https://www.science.org/doi/pdf/10.1126/sciadv.abh1542} \BibitemShut
  {NoStop}%
\bibitem [{\citenamefont {Zhu}\ \emph {et~al.}(2021)\citenamefont {Zhu},
  \citenamefont {Hochstetter}, \citenamefont {Loeffler}, \citenamefont
  {{Diaz-Alvarez}}, \citenamefont {Nakayama}, \citenamefont {Lizier},\ and\
  \citenamefont {Kuncic}}]{Zhu2021information}%
  \BibitemOpen
  \bibfield  {author} {\bibinfo {author} {\bibfnamefont {R.}~\bibnamefont
  {Zhu}}, \bibinfo {author} {\bibfnamefont {J.}~\bibnamefont {Hochstetter}},
  \bibinfo {author} {\bibfnamefont {A.}~\bibnamefont {Loeffler}}, \bibinfo
  {author} {\bibfnamefont {A.}~\bibnamefont {{Diaz-Alvarez}}}, \bibinfo
  {author} {\bibfnamefont {T.}~\bibnamefont {Nakayama}}, \bibinfo {author}
  {\bibfnamefont {J.~T.}\ \bibnamefont {Lizier}},\ and\ \bibinfo {author}
  {\bibfnamefont {Z.}~\bibnamefont {Kuncic}},\ }\bibfield  {title} {\bibinfo
  {title} {Information dynamics in neuromorphic nanowire networks},\ }\href
  {https://doi.org/10.1038/s41598-021-92170-7} {\bibfield  {journal} {\bibinfo
  {journal} {Scientific Reports}\ }\textbf {\bibinfo {volume} {11}},\ \bibinfo
  {pages} {13047} (\bibinfo {year} {2021})}\BibitemShut {NoStop}%
\bibitem [{\citenamefont {Caravelli}\ \emph
  {et~al.}(2023{\natexlab{a}})\citenamefont {Caravelli}, \citenamefont
  {Traversa}, \citenamefont {Bonnin},\ and\ \citenamefont
  {Bonani}}]{caravelli2022}%
  \BibitemOpen
  \bibfield  {author} {\bibinfo {author} {\bibfnamefont {F.}~\bibnamefont
  {Caravelli}}, \bibinfo {author} {\bibfnamefont {F.}~\bibnamefont {Traversa}},
  \bibinfo {author} {\bibfnamefont {M.}~\bibnamefont {Bonnin}},\ and\ \bibinfo
  {author} {\bibfnamefont {F.}~\bibnamefont {Bonani}},\ }\bibfield  {title}
  {\bibinfo {title} {Projective embedding of dynamical systems: Uniform mean
  field equations},\ }\href
  {https://doi.org/https://doi.org/10.1016/j.physd.2023.133747} {\bibfield
  {journal} {\bibinfo  {journal} {Physica D: Nonlinear Phenomena}\ }\textbf
  {\bibinfo {volume} {450}},\ \bibinfo {pages} {133747} (\bibinfo {year}
  {2023}{\natexlab{a}})}\BibitemShut {NoStop}%
\bibitem [{\citenamefont {Palmer}(1982)}]{palmer}%
  \BibitemOpen
  \bibfield  {author} {\bibinfo {author} {\bibfnamefont {R.}~\bibnamefont
  {Palmer}},\ }\bibfield  {title} {\bibinfo {title} {Broken ergodicity},\
  }\href {https://doi.org/10.1080/00018738200101438} {\bibfield  {journal}
  {\bibinfo  {journal} {Advances in Physics}\ }\textbf {\bibinfo {volume}
  {31}},\ \bibinfo {pages} {669} (\bibinfo {year} {1982})},\ \Eprint
  {https://arxiv.org/abs/https://doi.org/10.1080/00018738200101438}
  {https://doi.org/10.1080/00018738200101438} \BibitemShut {NoStop}%
\bibitem [{\citenamefont {Jaeger}\ and\ \citenamefont
  {Haas}(2004)}]{Jaeger2004}%
  \BibitemOpen
  \bibfield  {author} {\bibinfo {author} {\bibfnamefont {H.}~\bibnamefont
  {Jaeger}}\ and\ \bibinfo {author} {\bibfnamefont {H.}~\bibnamefont {Haas}},\
  }\bibfield  {title} {\bibinfo {title} {Harnessing nonlinearity: Predicting
  chaotic systems and saving energy in wireless communication},\ }\href
  {https://doi.org/10.1126/science.1091277} {\bibfield  {journal} {\bibinfo
  {journal} {Science}\ }\textbf {\bibinfo {volume} {304}},\ \bibinfo {pages}
  {78} (\bibinfo {year} {2004})}\BibitemShut {NoStop}%
\bibitem [{\citenamefont {Grigoryeva}\ and\ \citenamefont
  {Ortega}(2018)}]{Grigoryeva2018}%
  \BibitemOpen
  \bibfield  {author} {\bibinfo {author} {\bibfnamefont {L.}~\bibnamefont
  {Grigoryeva}}\ and\ \bibinfo {author} {\bibfnamefont {J.-P.}\ \bibnamefont
  {Ortega}},\ }\bibfield  {title} {\bibinfo {title} {Echo state networks are
  universal},\ }\href {https://doi.org/10.1016/j.neunet.2018.08.025} {\bibfield
   {journal} {\bibinfo  {journal} {Neural Networks}\ }\textbf {\bibinfo
  {volume} {108}},\ \bibinfo {pages} {495} (\bibinfo {year}
  {2018})}\BibitemShut {NoStop}%
\bibitem [{\citenamefont {Carroll}(2020)}]{Carroll2020}%
  \BibitemOpen
  \bibfield  {author} {\bibinfo {author} {\bibfnamefont {T.~L.}\ \bibnamefont
  {Carroll}},\ }\bibfield  {title} {\bibinfo {title} {Do reservoir computers
  work best at the edge of chaos?},\ }\bibfield  {journal} {\bibinfo  {journal}
  {Chaos: An Interdisciplinary Journal of Nonlinear Science}\ }\textbf
  {\bibinfo {volume} {30}},\ \href {https://doi.org/10.1063/5.0038163}
  {10.1063/5.0038163} (\bibinfo {year} {2020}),\ \bibinfo {note} {121109},\
  \Eprint
  {https://arxiv.org/abs/https://pubs.aip.org/aip/cha/article-pdf/doi/10.1063/5.0038163/14106721/121109\_1\_online.pdf}
  {https://pubs.aip.org/aip/cha/article-pdf/doi/10.1063/5.0038163/14106721/121109\_1\_online.pdf}
  \BibitemShut {NoStop}%
\bibitem [{\citenamefont {Tagliazucchi}\ \emph {et~al.}(2012)\citenamefont
  {Tagliazucchi}, \citenamefont {Balenzuela}, \citenamefont {Fraiman},\ and\
  \citenamefont {Chialvo}}]{Tagliazucchi2012}%
  \BibitemOpen
  \bibfield  {author} {\bibinfo {author} {\bibfnamefont {E.}~\bibnamefont
  {Tagliazucchi}}, \bibinfo {author} {\bibfnamefont {P.}~\bibnamefont
  {Balenzuela}}, \bibinfo {author} {\bibfnamefont {D.}~\bibnamefont
  {Fraiman}},\ and\ \bibinfo {author} {\bibfnamefont {D.~R.}\ \bibnamefont
  {Chialvo}},\ }\bibfield  {title} {\bibinfo {title} {Criticality in
  large-scale brain {fMRI} dynamics unveiled by a novel point process
  analysis},\ }\bibfield  {journal} {\bibinfo  {journal} {Frontiers in
  Physiology}\ }\textbf {\bibinfo {volume} {3}},\ \href
  {https://doi.org/10.3389/fphys.2012.00015} {10.3389/fphys.2012.00015}
  (\bibinfo {year} {2012})\BibitemShut {NoStop}%
\bibitem [{\citenamefont {Morales}\ and\ \citenamefont
  {Mu{\~{n}}oz}(2021)}]{Morales2021}%
  \BibitemOpen
  \bibfield  {author} {\bibinfo {author} {\bibfnamefont {G.~B.}\ \bibnamefont
  {Morales}}\ and\ \bibinfo {author} {\bibfnamefont {M.~A.}\ \bibnamefont
  {Mu{\~{n}}oz}},\ }\bibfield  {title} {\bibinfo {title} {Optimal input
  representation in neural systems at the edge of chaos},\ }\href
  {https://doi.org/10.3390/biology10080702} {\bibfield  {journal} {\bibinfo
  {journal} {Biology}\ }\textbf {\bibinfo {volume} {10}},\ \bibinfo {pages}
  {702} (\bibinfo {year} {2021})}\BibitemShut {NoStop}%
\bibitem [{\citenamefont {Vincent}(2007)}]{Vincent2007}%
  \BibitemOpen
  \bibfield  {author} {\bibinfo {author} {\bibfnamefont {E.}~\bibnamefont
  {Vincent}},\ }\bibinfo {title} {Ageing, rejuvenation and memory: The example
  of spin-glasses},\ in\ \href {https://doi.org/10.1007/3-540-69684-9_2} {\emph
  {\bibinfo {booktitle} {Ageing and the Glass Transition}}},\ \bibinfo {editor}
  {edited by\ \bibinfo {editor} {\bibfnamefont {M.}~\bibnamefont {Henkel}},
  \bibinfo {editor} {\bibfnamefont {M.}~\bibnamefont {Pleimling}},\ and\
  \bibinfo {editor} {\bibfnamefont {R.}~\bibnamefont {Sanctuary}}}\ (\bibinfo
  {publisher} {Springer Berlin Heidelberg},\ \bibinfo {address} {Berlin,
  Heidelberg},\ \bibinfo {year} {2007})\ pp.\ \bibinfo {pages}
  {7--60}\BibitemShut {NoStop}%
\bibitem [{\citenamefont {Saccone}\ \emph {et~al.}(2022)\citenamefont
  {Saccone}, \citenamefont {Caravelli}, \citenamefont {Hofhuis}, \citenamefont
  {Parchenko}, \citenamefont {Birkh\"{o}lzer}, \citenamefont {Dhuey},
  \citenamefont {Kleibert}, \citenamefont {van Dijken}, \citenamefont
  {Nisoli},\ and\ \citenamefont {Farhan}}]{Saccone2022}%
  \BibitemOpen
  \bibfield  {author} {\bibinfo {author} {\bibfnamefont {M.}~\bibnamefont
  {Saccone}}, \bibinfo {author} {\bibfnamefont {F.}~\bibnamefont {Caravelli}},
  \bibinfo {author} {\bibfnamefont {K.}~\bibnamefont {Hofhuis}}, \bibinfo
  {author} {\bibfnamefont {S.}~\bibnamefont {Parchenko}}, \bibinfo {author}
  {\bibfnamefont {Y.~A.}\ \bibnamefont {Birkh\"{o}lzer}}, \bibinfo {author}
  {\bibfnamefont {S.}~\bibnamefont {Dhuey}}, \bibinfo {author} {\bibfnamefont
  {A.}~\bibnamefont {Kleibert}}, \bibinfo {author} {\bibfnamefont
  {S.}~\bibnamefont {van Dijken}}, \bibinfo {author} {\bibfnamefont
  {C.}~\bibnamefont {Nisoli}},\ and\ \bibinfo {author} {\bibfnamefont
  {A.}~\bibnamefont {Farhan}},\ }\bibfield  {title} {\bibinfo {title} {Direct
  observation of a dynamical glass transition in a nanomagnetic artificial
  hopfield network},\ }\href {https://doi.org/10.1038/s41567-022-01538-7}
  {\bibfield  {journal} {\bibinfo  {journal} {Nature Physics}\ }\textbf
  {\bibinfo {volume} {18}},\ \bibinfo {pages} {517} (\bibinfo {year}
  {2022})}\BibitemShut {NoStop}%
\bibitem [{\citenamefont {Saccone}\ \emph {et~al.}(2023)\citenamefont
  {Saccone}, \citenamefont {Caravelli},\ and\ \citenamefont {al}}]{saccone}%
  \BibitemOpen
  \bibfield  {author} {\bibinfo {author} {\bibfnamefont {M.}~\bibnamefont
  {Saccone}}, \bibinfo {author} {\bibfnamefont {F.}~\bibnamefont {Caravelli}},\
  and\ \bibinfo {author} {\bibfnamefont {e.}~\bibnamefont {al}},\ }\bibfield
  {title} {\bibinfo {title} {Real-space observation of ergodicity transitions
  in artificial spin ice},\ }\href@noop {} {\bibfield  {journal} {\bibinfo
  {journal} {Nature Communications}\ }\textbf {\bibinfo {volume} {14}}
  (\bibinfo {year} {2023})}\BibitemShut {NoStop}%
\bibitem [{\citenamefont {Dorfman}(1999)}]{dorfmann99a}%
  \BibitemOpen
  \bibfield  {author} {\bibinfo {author} {\bibfnamefont {J.~R.}\ \bibnamefont
  {Dorfman}},\ }\href@noop {} {\emph {\bibinfo {title} {An Introduction to
  Chaos in Nonequilibrium Statistical Mechanics}}}\ (\bibinfo  {publisher}
  {Cambridge university Press},\ \bibinfo {year} {1999})\BibitemShut {NoStop}%
\bibitem [{\citenamefont {Ma}(1985)}]{mastatistical}%
  \BibitemOpen
  \bibfield  {author} {\bibinfo {author} {\bibfnamefont {S.-K.}\ \bibnamefont
  {Ma}},\ }\href@noop {} {\emph {\bibinfo {title} {Statistical Mechanics}}}\
  (\bibinfo  {publisher} {World Scientific},\ \bibinfo {year}
  {1985})\BibitemShut {NoStop}%
\bibitem [{\citenamefont {Thirumalai}\ \emph {et~al.}(1989)\citenamefont
  {Thirumalai}, \citenamefont {Mountain},\ and\ \citenamefont
  {Kirkpatrick}}]{mountain89me}%
  \BibitemOpen
  \bibfield  {author} {\bibinfo {author} {\bibfnamefont {D.}~\bibnamefont
  {Thirumalai}}, \bibinfo {author} {\bibfnamefont {R.~D.}\ \bibnamefont
  {Mountain}},\ and\ \bibinfo {author} {\bibfnamefont {T.~R.}\ \bibnamefont
  {Kirkpatrick}},\ }\bibfield  {title} {\bibinfo {title} {Ergodic behavior in
  supercooled liquids and in glasses},\ }\href
  {https://doi.org/10.1103/PhysRevA.39.3563} {\bibfield  {journal} {\bibinfo
  {journal} {Phys. Rev. A}\ }\textbf {\bibinfo {volume} {39}},\ \bibinfo
  {pages} {3563} (\bibinfo {year} {1989})}\BibitemShut {NoStop}%
\bibitem [{\citenamefont {Petersen}(1983)}]{ergodictheory}%
  \BibitemOpen
  \bibfield  {author} {\bibinfo {author} {\bibfnamefont {K.}~\bibnamefont
  {Petersen}},\ }\href@noop {} {\emph {\bibinfo {title} {Ergodic theory}}}\
  (\bibinfo  {publisher} {World Scientific},\ \bibinfo {year}
  {1983})\BibitemShut {NoStop}%
\bibitem [{\citenamefont {Mountain}\ and\ \citenamefont
  {Thirumalai}(1989)}]{mountain_measures_1989}%
  \BibitemOpen
  \bibfield  {author} {\bibinfo {author} {\bibfnamefont {R.~D.}\ \bibnamefont
  {Mountain}}\ and\ \bibinfo {author} {\bibfnamefont {D.}~\bibnamefont
  {Thirumalai}},\ }\bibfield  {title} {\bibinfo {title} {Measures of effective
  ergodic convergence in liquids},\ }\href
  {https://doi.org/10.1021/j100356a019} {\bibfield  {journal} {\bibinfo
  {journal} {The Journal of Physical Chemistry}\ }\textbf {\bibinfo {volume}
  {93}},\ \bibinfo {pages} {6975} (\bibinfo {year} {1989})},\ \bibinfo {note}
  {publisher: American Chemical Society}\BibitemShut {NoStop}%
\bibitem [{\citenamefont {Thirumalai}\ and\ \citenamefont
  {Mountain}(1993)}]{mountain93me}%
  \BibitemOpen
  \bibfield  {author} {\bibinfo {author} {\bibfnamefont {D.}~\bibnamefont
  {Thirumalai}}\ and\ \bibinfo {author} {\bibfnamefont {R.~D.}\ \bibnamefont
  {Mountain}},\ }\bibfield  {title} {\bibinfo {title} {Activated dynamics, loss
  of ergodicity, and transport in supercooled liquids},\ }\href
  {https://doi.org/10.1103/PhysRevE.47.479} {\bibfield  {journal} {\bibinfo
  {journal} {Phys. Rev. E}\ }\textbf {\bibinfo {volume} {47}},\ \bibinfo
  {pages} {479} (\bibinfo {year} {1993})}\BibitemShut {NoStop}%
\bibitem [{\citenamefont {Bel}\ and\ \citenamefont {Barkai}(2006)}]{barkai}%
  \BibitemOpen
  \bibfield  {author} {\bibinfo {author} {\bibfnamefont {G.}~\bibnamefont
  {Bel}}\ and\ \bibinfo {author} {\bibfnamefont {E.}~\bibnamefont {Barkai}},\
  }\bibfield  {title} {\bibinfo {title} {Ergodicity breaking in a deterministic
  dynamical system},\ }\href@noop {} {\bibfield  {journal} {\bibinfo  {journal}
  {Europhys. Lett.}\ }\textbf {\bibinfo {volume} {74}},\ \bibinfo {pages} {15}
  (\bibinfo {year} {2006})}\BibitemShut {NoStop}%
\bibitem [{\citenamefont {Tiampo}\ \emph {et~al.}(2003)\citenamefont {Tiampo},
  \citenamefont {Rundle}, \citenamefont {Klein}, \citenamefont {Martins},\ and\
  \citenamefont {Ferguson}}]{tiampo03a}%
  \BibitemOpen
  \bibfield  {author} {\bibinfo {author} {\bibfnamefont {K.~F.}\ \bibnamefont
  {Tiampo}}, \bibinfo {author} {\bibfnamefont {J.~B.}\ \bibnamefont {Rundle}},
  \bibinfo {author} {\bibfnamefont {W.}~\bibnamefont {Klein}}, \bibinfo
  {author} {\bibfnamefont {J.~S.~S.}\ \bibnamefont {Martins}},\ and\ \bibinfo
  {author} {\bibfnamefont {C.~D.}\ \bibnamefont {Ferguson}},\ }\bibfield
  {title} {\bibinfo {title} {Ergodic dynamics in a natural threshold system},\
  }\href {https://doi.org/10.1103/PhysRevLett.91.238501} {\bibfield  {journal}
  {\bibinfo  {journal} {Phys. Rev. Lett.}\ }\textbf {\bibinfo {volume} {91}},\
  \bibinfo {pages} {238501} (\bibinfo {year} {2003})}\BibitemShut {NoStop}%
\bibitem [{\citenamefont {Tiampo}\ \emph {et~al.}(2007)\citenamefont {Tiampo},
  \citenamefont {Rundle}, \citenamefont {Klein}, \citenamefont {Holliday},
  \citenamefont {S\'a~Martins},\ and\ \citenamefont {Ferguson}}]{tiampo2007a}%
  \BibitemOpen
  \bibfield  {author} {\bibinfo {author} {\bibfnamefont {K.~F.}\ \bibnamefont
  {Tiampo}}, \bibinfo {author} {\bibfnamefont {J.~B.}\ \bibnamefont {Rundle}},
  \bibinfo {author} {\bibfnamefont {W.}~\bibnamefont {Klein}}, \bibinfo
  {author} {\bibfnamefont {J.}~\bibnamefont {Holliday}}, \bibinfo {author}
  {\bibfnamefont {J.~S.}\ \bibnamefont {S\'a~Martins}},\ and\ \bibinfo {author}
  {\bibfnamefont {C.~D.}\ \bibnamefont {Ferguson}},\ }\bibfield  {title}
  {\bibinfo {title} {Ergodicity in natural earthquake fault networks},\ }\href
  {https://doi.org/10.1103/PhysRevE.75.066107} {\bibfield  {journal} {\bibinfo
  {journal} {Phys. Rev. E}\ }\textbf {\bibinfo {volume} {75}},\ \bibinfo
  {pages} {066107} (\bibinfo {year} {2007})}\BibitemShut {NoStop}%
\bibitem [{\citenamefont {S\"uzen}(2014)}]{suzen}%
  \BibitemOpen
  \bibfield  {author} {\bibinfo {author} {\bibfnamefont {M.}~\bibnamefont
  {S\"uzen}},\ }\bibfield  {title} {\bibinfo {title} {Effective ergodicity in
  single-spin-flip dynamics},\ }\href
  {https://doi.org/10.1103/PhysRevE.90.032141} {\bibfield  {journal} {\bibinfo
  {journal} {Phys. Rev. E}\ }\textbf {\bibinfo {volume} {90}},\ \bibinfo
  {pages} {032141} (\bibinfo {year} {2014})}\BibitemShut {NoStop}%
\bibitem [{\citenamefont {Chua}(1971)}]{Chua1971}%
  \BibitemOpen
  \bibfield  {author} {\bibinfo {author} {\bibfnamefont {L.}~\bibnamefont
  {Chua}},\ }\bibfield  {title} {\bibinfo {title} {Memristor-the missing
  circuit element},\ }\href {https://doi.org/10.1109/TCT.1971.1083337}
  {\bibfield  {journal} {\bibinfo  {journal} {IEEE Transactions on Circuit
  Theory}\ }\textbf {\bibinfo {volume} {18}},\ \bibinfo {pages} {507} (\bibinfo
  {year} {1971})}\BibitemShut {NoStop}%
\bibitem [{\citenamefont {Caravelli}(2017{\natexlab{a}})}]{Caravelli2017b}%
  \BibitemOpen
  \bibfield  {author} {\bibinfo {author} {\bibfnamefont {F.}~\bibnamefont
  {Caravelli}},\ }\bibfield  {title} {\bibinfo {title} {Locality of
  interactions for planar memristive circuits},\ }\href
  {https://doi.org/10.1103/physreve.96.052206} {\bibfield  {journal} {\bibinfo
  {journal} {Phys. Rev. E}\ }\textbf {\bibinfo {volume} {96}},\ \bibinfo
  {pages} {052206} (\bibinfo {year} {2017}{\natexlab{a}})}\BibitemShut
  {NoStop}%
\bibitem [{\citenamefont {Joglekar}\ and\ \citenamefont
  {Wolf}(2009)}]{Joglekar2009}%
  \BibitemOpen
  \bibfield  {author} {\bibinfo {author} {\bibfnamefont {Y.~N.}\ \bibnamefont
  {Joglekar}}\ and\ \bibinfo {author} {\bibfnamefont {S.~J.}\ \bibnamefont
  {Wolf}},\ }\bibfield  {title} {\bibinfo {title} {The elusive memristor:
  properties of basic electrical circuits},\ }\href
  {https://doi.org/10.1088/0143-0807/30/4/001} {\bibfield  {journal} {\bibinfo
  {journal} {Eur. J. of Phys.}\ }\textbf {\bibinfo {volume} {30}},\ \bibinfo
  {pages} {661} (\bibinfo {year} {2009})}\BibitemShut {NoStop}%
\bibitem [{\citenamefont {Prodromakis}\ and\ \citenamefont {et.
  al.}(2011)}]{Prodromakis2011}%
  \BibitemOpen
  \bibfield  {author} {\bibinfo {author} {\bibfnamefont {T.}~\bibnamefont
  {Prodromakis}}\ and\ \bibinfo {author} {\bibnamefont {et. al.}},\ }\bibfield
  {title} {\bibinfo {title} {A versatile memristor model with nonlinear dopant
  kinetics},\ }\href {https://doi.org/10.1109/ted.2011.2158004} {\bibfield
  {journal} {\bibinfo  {journal} {{IEEE} Trans. on El. Dev.}\ }\textbf
  {\bibinfo {volume} {58}},\ \bibinfo {pages} {3099} (\bibinfo {year}
  {2011})}\BibitemShut {NoStop}%
\bibitem [{\citenamefont {Caravelli}\ \emph {et~al.}(2017)\citenamefont
  {Caravelli}, \citenamefont {Traversa},\ and\ \citenamefont {{Di
  Ventra}}}]{Caravelli2016rl}%
  \BibitemOpen
  \bibfield  {author} {\bibinfo {author} {\bibfnamefont {F.}~\bibnamefont
  {Caravelli}}, \bibinfo {author} {\bibfnamefont {F.~L.}\ \bibnamefont
  {Traversa}},\ and\ \bibinfo {author} {\bibfnamefont {M.}~\bibnamefont {{Di
  Ventra}}},\ }\bibfield  {title} {\bibinfo {title} {Complex dynamics of
  memristive circuits: Analytical results and universal slow relaxation},\
  }\href {https://doi.org/10.1103/physreve.95.022140} {\bibfield  {journal}
  {\bibinfo  {journal} {Phys. Rev. E}\ }\textbf {\bibinfo {volume} {95}},\
  \bibinfo {pages} {022140} (\bibinfo {year} {2017})}\BibitemShut {NoStop}%
\bibitem [{\citenamefont {Li}\ \emph {et~al.}(2020)\citenamefont {Li},
  \citenamefont {Diaz-Alvarez}, \citenamefont {Iguchi}, \citenamefont
  {Hochstetter}, \citenamefont {Loeffler}, \citenamefont {Zhu}, \citenamefont
  {Shingaya}, \citenamefont {Kuncic}, \citenamefont {Uchida},\ and\
  \citenamefont {Nakayama}}]{Li2020}%
  \BibitemOpen
  \bibfield  {author} {\bibinfo {author} {\bibfnamefont {Q.}~\bibnamefont
  {Li}}, \bibinfo {author} {\bibfnamefont {A.}~\bibnamefont {Diaz-Alvarez}},
  \bibinfo {author} {\bibfnamefont {R.}~\bibnamefont {Iguchi}}, \bibinfo
  {author} {\bibfnamefont {J.}~\bibnamefont {Hochstetter}}, \bibinfo {author}
  {\bibfnamefont {A.}~\bibnamefont {Loeffler}}, \bibinfo {author}
  {\bibfnamefont {R.}~\bibnamefont {Zhu}}, \bibinfo {author} {\bibfnamefont
  {Y.}~\bibnamefont {Shingaya}}, \bibinfo {author} {\bibfnamefont
  {Z.}~\bibnamefont {Kuncic}}, \bibinfo {author} {\bibfnamefont {K.-i.}\
  \bibnamefont {Uchida}},\ and\ \bibinfo {author} {\bibfnamefont
  {T.}~\bibnamefont {Nakayama}},\ }\bibfield  {title} {\bibinfo {title}
  {Dynamic electrical pathway tuning in neuromorphic nanowire networks},\
  }\href {https://doi.org/https://doi.org/10.1002/adfm.202003679} {\bibfield
  {journal} {\bibinfo  {journal} {Advanced Functional Materials}\ }\textbf
  {\bibinfo {volume} {30}},\ \bibinfo {pages} {2003679} (\bibinfo {year}
  {2020})}\BibitemShut {NoStop}%
\bibitem [{\citenamefont {Loeffler}\ \emph {et~al.}(2020)\citenamefont
  {Loeffler}, \citenamefont {Zhu}, \citenamefont {Hochstetter}, \citenamefont
  {Li}, \citenamefont {Fu}, \citenamefont {Diaz-Alvarez}, \citenamefont
  {Nakayama}, \citenamefont {Shine},\ and\ \citenamefont
  {Kuncic}}]{Loeffler2020}%
  \BibitemOpen
  \bibfield  {author} {\bibinfo {author} {\bibfnamefont {A.}~\bibnamefont
  {Loeffler}}, \bibinfo {author} {\bibfnamefont {R.}~\bibnamefont {Zhu}},
  \bibinfo {author} {\bibfnamefont {J.}~\bibnamefont {Hochstetter}}, \bibinfo
  {author} {\bibfnamefont {M.}~\bibnamefont {Li}}, \bibinfo {author}
  {\bibfnamefont {K.}~\bibnamefont {Fu}}, \bibinfo {author} {\bibfnamefont
  {A.}~\bibnamefont {Diaz-Alvarez}}, \bibinfo {author} {\bibfnamefont
  {T.}~\bibnamefont {Nakayama}}, \bibinfo {author} {\bibfnamefont {J.~M.}\
  \bibnamefont {Shine}},\ and\ \bibinfo {author} {\bibfnamefont
  {Z.}~\bibnamefont {Kuncic}},\ }\bibfield  {title} {\bibinfo {title}
  {Topological properties of neuromorphic nanowire networks},\ }\bibfield
  {journal} {\bibinfo  {journal} {Frontiers in Neuroscience}\ }\textbf
  {\bibinfo {volume} {14}},\ \href {https://doi.org/10.3389/fnins.2020.00184}
  {10.3389/fnins.2020.00184} (\bibinfo {year} {2020})\BibitemShut {NoStop}%
\bibitem [{\citenamefont {Kuncic}\ \emph {et~al.}(2020)\citenamefont {Kuncic},
  \citenamefont {Kavehei}, \citenamefont {Zhu}, \citenamefont {Loeffler},
  \citenamefont {Fu}, \citenamefont {Hochstetter}, \citenamefont {Li},
  \citenamefont {Shine}, \citenamefont {{Diaz-Alvarez}}, \citenamefont {Stieg},
  \citenamefont {Gimzewski},\ and\ \citenamefont {Nakayama}}]{Kuncic2020}%
  \BibitemOpen
  \bibfield  {author} {\bibinfo {author} {\bibfnamefont {Z.}~\bibnamefont
  {Kuncic}}, \bibinfo {author} {\bibfnamefont {O.}~\bibnamefont {Kavehei}},
  \bibinfo {author} {\bibfnamefont {R.}~\bibnamefont {Zhu}}, \bibinfo {author}
  {\bibfnamefont {A.}~\bibnamefont {Loeffler}}, \bibinfo {author}
  {\bibfnamefont {K.}~\bibnamefont {Fu}}, \bibinfo {author} {\bibfnamefont
  {J.}~\bibnamefont {Hochstetter}}, \bibinfo {author} {\bibfnamefont
  {M.}~\bibnamefont {Li}}, \bibinfo {author} {\bibfnamefont {J.~M.}\
  \bibnamefont {Shine}}, \bibinfo {author} {\bibfnamefont {A.}~\bibnamefont
  {{Diaz-Alvarez}}}, \bibinfo {author} {\bibfnamefont {A.}~\bibnamefont
  {Stieg}}, \bibinfo {author} {\bibfnamefont {J.}~\bibnamefont {Gimzewski}},\
  and\ \bibinfo {author} {\bibfnamefont {T.}~\bibnamefont {Nakayama}},\
  }\bibfield  {title} {\bibinfo {title} {Neuromorphic {{Information
  Processing}} with {{Nanowire Networks}}},\ }in\ \href
  {https://doi.org/10.1109/ISCAS45731.2020.9181034} {\emph {\bibinfo
  {booktitle} {2020 {{IEEE International Symposium}} on {{Circuits}} and
  {{Systems}} ({{ISCAS}})}}}\ (\bibinfo {year} {2020})\ pp.\ \bibinfo {pages}
  {1--5}\BibitemShut {NoStop}%
\bibitem [{\citenamefont {Nilsson}\ and\ \citenamefont
  {Riedel}(2011)}]{Nilsson}%
  \BibitemOpen
  \bibfield  {author} {\bibinfo {author} {\bibfnamefont {J.~W.}\ \bibnamefont
  {Nilsson}}\ and\ \bibinfo {author} {\bibfnamefont {S.}~\bibnamefont
  {Riedel}},\ }\href {https://doi.org/: 0-13-376003-0} {\emph {\bibinfo {title}
  {Electric Circuits (9th ed),}}}\ (\bibinfo  {publisher} {Pearson Education,
  Saddle River NJ},\ \bibinfo {year} {2011})\BibitemShut {NoStop}%
\bibitem [{\citenamefont {Fu}\ \emph {et~al.}(2020)\citenamefont {Fu},
  \citenamefont {Zhu}, \citenamefont {Loeffler},\ and\ \citenamefont
  {al}}]{zhurc}%
  \BibitemOpen
  \bibfield  {author} {\bibinfo {author} {\bibfnamefont {K.}~\bibnamefont
  {Fu}}, \bibinfo {author} {\bibfnamefont {R.}~\bibnamefont {Zhu}}, \bibinfo
  {author} {\bibfnamefont {A.}~\bibnamefont {Loeffler}},\ and\ \bibinfo
  {author} {\bibfnamefont {e.}~\bibnamefont {al}},\ }\bibfield  {title}
  {\bibinfo {title} {Reservoir computing with neuromemristive nanowire
  networks},\ }\href@noop {} {\bibfield  {journal} {\bibinfo  {journal}
  {Proceedings of the International Joint Conference on Neural Networks
  (IJCNN)}\ }\textbf {\bibinfo {volume} {20006228}} (\bibinfo {year}
  {2020})}\BibitemShut {NoStop}%
\bibitem [{\citenamefont {Caravelli}\ and\ \citenamefont
  {Sheldon}(2020)}]{caravellisheldon}%
  \BibitemOpen
  \bibfield  {author} {\bibinfo {author} {\bibfnamefont {F.}~\bibnamefont
  {Caravelli}}\ and\ \bibinfo {author} {\bibfnamefont {F.~C.}\ \bibnamefont
  {Sheldon}},\ }\bibfield  {title} {\bibinfo {title} {Phases of memristive
  circuits via an interacting disorder approach},\ }\href@noop {} {\bibfield
  {journal} {\bibinfo  {journal} {https://arxiv.org/abs/2009.00114}\ }
  (\bibinfo {year} {2020})}\BibitemShut {NoStop}%
\bibitem [{\citenamefont {Sheldon}\ \emph {et~al.}(2022)\citenamefont
  {Sheldon}, \citenamefont {Kolchinsky},\ and\ \citenamefont
  {Caravelli}}]{sheldonrc}%
  \BibitemOpen
  \bibfield  {author} {\bibinfo {author} {\bibfnamefont {F.~C.}\ \bibnamefont
  {Sheldon}}, \bibinfo {author} {\bibfnamefont {A.}~\bibnamefont
  {Kolchinsky}},\ and\ \bibinfo {author} {\bibfnamefont {F.}~\bibnamefont
  {Caravelli}},\ }\bibfield  {title} {\bibinfo {title} {The computational
  capacity of lrc, memristive and hybrid reservoirs},\ }\href@noop {}
  {\bibfield  {journal} {\bibinfo  {journal} {Phys. Rev. E}\ }\textbf {\bibinfo
  {volume} {106}} (\bibinfo {year} {2022})}\BibitemShut {NoStop}%
\bibitem [{\citenamefont {Caravelli}\ \emph
  {et~al.}(2023{\natexlab{b}})\citenamefont {Caravelli}, \citenamefont
  {Milano}, \citenamefont {Ricciardi},\ and\ \citenamefont
  {Kuncic}}]{Caravelli2023}%
  \BibitemOpen
  \bibfield  {author} {\bibinfo {author} {\bibfnamefont {F.}~\bibnamefont
  {Caravelli}}, \bibinfo {author} {\bibfnamefont {G.}~\bibnamefont {Milano}},
  \bibinfo {author} {\bibfnamefont {C.}~\bibnamefont {Ricciardi}},\ and\
  \bibinfo {author} {\bibfnamefont {Z.}~\bibnamefont {Kuncic}},\ }\bibfield
  {title} {\bibinfo {title} {Mean field theory of self-organizing memristive
  connectomes},\ }\bibfield  {journal} {\bibinfo  {journal} {Annalen der
  Physik}\ }\textbf {\bibinfo {volume} {535}},\ \href
  {https://doi.org/10.1002/andp.202300090} {10.1002/andp.202300090} (\bibinfo
  {year} {2023}{\natexlab{b}})\BibitemShut {NoStop}%
\bibitem [{\citenamefont {Manning}\ \emph {et~al.}(2018)\citenamefont
  {Manning}, \citenamefont {Niosi}, \citenamefont {da~Rocha},\ and\
  \citenamefont {al.}}]{wta}%
  \BibitemOpen
  \bibfield  {author} {\bibinfo {author} {\bibfnamefont {H.}~\bibnamefont
  {Manning}}, \bibinfo {author} {\bibfnamefont {F.}~\bibnamefont {Niosi}},
  \bibinfo {author} {\bibfnamefont {C.}~\bibnamefont {da~Rocha}},\ and\
  \bibinfo {author} {\bibfnamefont {e.}~\bibnamefont {al.}},\ }\bibfield
  {title} {\bibinfo {title} {Emergence of winner-takes-all connectivity paths
  in random nanowire networks},\ }\href@noop {} {\bibfield  {journal} {\bibinfo
   {journal} {Nat Commun}\ }\textbf {\bibinfo {volume} {9}} (\bibinfo {year}
  {2018})}\BibitemShut {NoStop}%
\bibitem [{\citenamefont {Daniels}\ \emph {et~al.}(2022)\citenamefont
  {Daniels}, \citenamefont {Mallinson}, \citenamefont {Heywood}, \citenamefont
  {Bones}, \citenamefont {Arnold},\ and\ \citenamefont {Brown}}]{Daniels2022}%
  \BibitemOpen
  \bibfield  {author} {\bibinfo {author} {\bibfnamefont {R.}~\bibnamefont
  {Daniels}}, \bibinfo {author} {\bibfnamefont {J.}~\bibnamefont {Mallinson}},
  \bibinfo {author} {\bibfnamefont {Z.}~\bibnamefont {Heywood}}, \bibinfo
  {author} {\bibfnamefont {P.}~\bibnamefont {Bones}}, \bibinfo {author}
  {\bibfnamefont {M.}~\bibnamefont {Arnold}},\ and\ \bibinfo {author}
  {\bibfnamefont {S.}~\bibnamefont {Brown}},\ }\bibfield  {title} {\bibinfo
  {title} {Reservoir computing with {{3D}} nanowire networks},\ }\href
  {https://doi.org/10.1016/j.neunet.2022.07.001} {\bibfield  {journal}
  {\bibinfo  {journal} {Neural Networks}\ }\textbf {\bibinfo {volume} {154}},\
  \bibinfo {pages} {122} (\bibinfo {year} {2022})}\BibitemShut {NoStop}%
\bibitem [{\citenamefont {Appeltant}\ \emph {et~al.}(2011)\citenamefont
  {Appeltant}, \citenamefont {Soriano}, \citenamefont {der Sande},
  \citenamefont {Danckaert}, \citenamefont {Massar}, \citenamefont {Dambre},
  \citenamefont {Schrauwen}, \citenamefont {Mirasso},\ and\ \citenamefont
  {Fischer}}]{Appeltant2011}%
  \BibitemOpen
  \bibfield  {author} {\bibinfo {author} {\bibfnamefont {L.}~\bibnamefont
  {Appeltant}}, \bibinfo {author} {\bibfnamefont {M.}~\bibnamefont {Soriano}},
  \bibinfo {author} {\bibfnamefont {G.~V.}\ \bibnamefont {der Sande}}, \bibinfo
  {author} {\bibfnamefont {J.}~\bibnamefont {Danckaert}}, \bibinfo {author}
  {\bibfnamefont {S.}~\bibnamefont {Massar}}, \bibinfo {author} {\bibfnamefont
  {J.}~\bibnamefont {Dambre}}, \bibinfo {author} {\bibfnamefont
  {B.}~\bibnamefont {Schrauwen}}, \bibinfo {author} {\bibfnamefont
  {C.}~\bibnamefont {Mirasso}},\ and\ \bibinfo {author} {\bibfnamefont
  {I.}~\bibnamefont {Fischer}},\ }\bibfield  {title} {\bibinfo {title}
  {Information processing using a single dynamical node as complex system},\
  }\bibfield  {journal} {\bibinfo  {journal} {Nature Communications}\ }\textbf
  {\bibinfo {volume} {2}},\ \href {https://doi.org/10.1038/ncomms1476}
  {10.1038/ncomms1476} (\bibinfo {year} {2011})\BibitemShut {NoStop}%
\bibitem [{\citenamefont {Caravelli}(2019)}]{Caravelli2019Ent}%
  \BibitemOpen
  \bibfield  {author} {\bibinfo {author} {\bibfnamefont {F.}~\bibnamefont
  {Caravelli}},\ }\bibfield  {title} {\bibinfo {title} {Asymptotic behavior of
  memristive circuits},\ }\href {https://doi.org/10.3390/e21080789} {\bibfield
  {journal} {\bibinfo  {journal} {Entropy}\ }\textbf {\bibinfo {volume} {21}},\
  \bibinfo {pages} {789} (\bibinfo {year} {2019})}\BibitemShut {NoStop}%
\bibitem [{\citenamefont {Zhu}\ \emph {et~al.}(2020)\citenamefont {Zhu},
  \citenamefont {Hochstetter}, \citenamefont {Loeffler}, \citenamefont
  {Diaz-Alvarez}, \citenamefont {Stieg}, \citenamefont {Gimzewski},
  \citenamefont {Nakayama},\ and\ \citenamefont {Kuncic}}]{loeffler}%
  \BibitemOpen
  \bibfield  {author} {\bibinfo {author} {\bibfnamefont {R.}~\bibnamefont
  {Zhu}}, \bibinfo {author} {\bibfnamefont {J.}~\bibnamefont {Hochstetter}},
  \bibinfo {author} {\bibfnamefont {A.}~\bibnamefont {Loeffler}}, \bibinfo
  {author} {\bibfnamefont {A.}~\bibnamefont {Diaz-Alvarez}}, \bibinfo {author}
  {\bibfnamefont {A.}~\bibnamefont {Stieg}}, \bibinfo {author} {\bibfnamefont
  {J.}~\bibnamefont {Gimzewski}}, \bibinfo {author} {\bibfnamefont
  {T.}~\bibnamefont {Nakayama}},\ and\ \bibinfo {author} {\bibfnamefont
  {Z.}~\bibnamefont {Kuncic}},\ }\bibfield  {title} {\bibinfo {title}
  {Harnessing adaptive dynamics in neuro-memristive nanowire networks for
  transfer learning},\ }in\ \href {https://doi.org/10.1109/ICRC2020.2020.00007}
  {\emph {\bibinfo {booktitle} {2020 International Conference on Rebooting
  Computing (ICRC)}}}\ (\bibinfo {year} {2020})\ pp.\ \bibinfo {pages}
  {102--106}\BibitemShut {NoStop}%
\bibitem [{\citenamefont {Milano}\ \emph {et~al.}(2022)\citenamefont {Milano},
  \citenamefont {Pedretti}, \citenamefont {Montano},\ and\ \citenamefont
  {al.}}]{milano001}%
  \BibitemOpen
  \bibfield  {author} {\bibinfo {author} {\bibfnamefont {G.}~\bibnamefont
  {Milano}}, \bibinfo {author} {\bibfnamefont {G.}~\bibnamefont {Pedretti}},
  \bibinfo {author} {\bibfnamefont {K.}~\bibnamefont {Montano}},\ and\ \bibinfo
  {author} {\bibfnamefont {e.}~\bibnamefont {al.}},\ }\bibfield  {title}
  {\bibinfo {title} {In materia reservoir computing with a fully memristive
  architecture based on self-organizing nanowire networks},\ }\href@noop {}
  {\bibfield  {journal} {\bibinfo  {journal} {Nat. Mater.}\ ,\ \bibinfo {pages}
  {195–202}} (\bibinfo {year} {2022})}\BibitemShut {NoStop}%
\bibitem [{\citenamefont {Loeffler}\ \emph {et~al.}(2023)\citenamefont
  {Loeffler}, \citenamefont {Diaz-Alvarez}, \citenamefont {Zhu}, \citenamefont
  {Ganesh}, \citenamefont {Shine}, \citenamefont {Nakayama},\ and\
  \citenamefont {Kuncic}}]{Loeffler2023}%
  \BibitemOpen
  \bibfield  {author} {\bibinfo {author} {\bibfnamefont {A.}~\bibnamefont
  {Loeffler}}, \bibinfo {author} {\bibfnamefont {A.}~\bibnamefont
  {Diaz-Alvarez}}, \bibinfo {author} {\bibfnamefont {R.}~\bibnamefont {Zhu}},
  \bibinfo {author} {\bibfnamefont {N.}~\bibnamefont {Ganesh}}, \bibinfo
  {author} {\bibfnamefont {J.~M.}\ \bibnamefont {Shine}}, \bibinfo {author}
  {\bibfnamefont {T.}~\bibnamefont {Nakayama}},\ and\ \bibinfo {author}
  {\bibfnamefont {Z.}~\bibnamefont {Kuncic}},\ }\bibfield  {title} {\bibinfo
  {title} {Neuromorphic learning, working memory, and metaplasticity in
  nanowire networks},\ }\bibfield  {journal} {\bibinfo  {journal} {Science
  Advances}\ }\textbf {\bibinfo {volume} {9}},\ \href
  {https://doi.org/10.1126/sciadv.adg3289} {10.1126/sciadv.adg3289} (\bibinfo
  {year} {2023})\BibitemShut {NoStop}%
\bibitem [{\citenamefont {Bartolucci}\ \emph {et~al.}(2020)\citenamefont
  {Bartolucci}, \citenamefont {Caccioli}, \citenamefont {Caravelli},\ and\
  \citenamefont {Vivo}}]{BCCV}%
  \BibitemOpen
  \bibfield  {author} {\bibinfo {author} {\bibfnamefont {S.}~\bibnamefont
  {Bartolucci}}, \bibinfo {author} {\bibfnamefont {F.}~\bibnamefont
  {Caccioli}}, \bibinfo {author} {\bibfnamefont {F.}~\bibnamefont
  {Caravelli}},\ and\ \bibinfo {author} {\bibfnamefont {P.}~\bibnamefont
  {Vivo}},\ }\bibfield  {title} {\bibinfo {title} {Inversion-free leontief
  inverse: statistical regularities in input-output analysis from partial
  information},\ }\href@noop {} {\bibfield  {journal} {\bibinfo  {journal}
  {https://arxiv.org/abs/2009.06350}\ } (\bibinfo {year} {2020})}\BibitemShut
  {NoStop}%
\bibitem [{\citenamefont {Bartolucci}\ \emph {et~al.}(2023)\citenamefont
  {Bartolucci}, \citenamefont {Caccioli}, \citenamefont {Caravelli},\ and\
  \citenamefont {Vivo}}]{BCCV2}%
  \BibitemOpen
  \bibfield  {author} {\bibinfo {author} {\bibfnamefont {S.}~\bibnamefont
  {Bartolucci}}, \bibinfo {author} {\bibfnamefont {F.}~\bibnamefont
  {Caccioli}}, \bibinfo {author} {\bibfnamefont {F.}~\bibnamefont
  {Caravelli}},\ and\ \bibinfo {author} {\bibfnamefont {P.}~\bibnamefont
  {Vivo}},\ }\bibfield  {title} {\bibinfo {title} {Ranking influential nodes in
  networks from aggregate local information},\ }\bibfield  {journal} {\bibinfo
  {journal} {Physical Review Research}\ }\textbf {\bibinfo {volume} {5}},\
  \href {https://doi.org/10.1103/physrevresearch.5.033123}
  {10.1103/physrevresearch.5.033123} (\bibinfo {year} {2023})\BibitemShut
  {NoStop}%
\bibitem [{\citenamefont {Caravelli}(2023)}]{caravelliwein}%
  \BibitemOpen
  \bibfield  {author} {\bibinfo {author} {\bibfnamefont {F.}~\bibnamefont
  {Caravelli}},\ }\bibfield  {title} {\bibinfo {title} {Cycle equivalence
  classes, orthogonal weingarten calculus, and the mean field theory of
  memristive systems},\ }\href@noop {} {\bibfield  {journal} {\bibinfo
  {journal} {arXiv:2304.14890}\ } (\bibinfo {year} {2023})}\BibitemShut
  {NoStop}%
\bibitem [{\citenamefont {Langton}(1990)}]{edgeofchaos}%
  \BibitemOpen
  \bibfield  {author} {\bibinfo {author} {\bibfnamefont {C.~G.}\ \bibnamefont
  {Langton}},\ }\bibfield  {title} {\bibinfo {title} {Computation at the edge
  of chaos: phase transitions and emergent computation},\ }\href@noop {}
  {\bibfield  {journal} {\bibinfo  {journal} {Physica D}\ }\textbf {\bibinfo
  {volume} {42}},\ \bibinfo {pages} {12} (\bibinfo {year} {1990})}\BibitemShut
  {NoStop}%
\bibitem [{\citenamefont {Chialvo}\ and\ \citenamefont
  {Bak}(1999)}]{Chialvo1999}%
  \BibitemOpen
  \bibfield  {author} {\bibinfo {author} {\bibfnamefont {D.}~\bibnamefont
  {Chialvo}}\ and\ \bibinfo {author} {\bibfnamefont {P.}~\bibnamefont {Bak}},\
  }\bibfield  {title} {\bibinfo {title} {Learning from mistakes},\ }\href
  {https://doi.org/10.1016/s0306-4522(98)00472-2} {\bibfield  {journal}
  {\bibinfo  {journal} {Neuroscience}\ }\textbf {\bibinfo {volume} {90}},\
  \bibinfo {pages} {1137} (\bibinfo {year} {1999})}\BibitemShut {NoStop}%
\bibitem [{\citenamefont {Carbajal}\ \emph {et~al.}(2022)\citenamefont
  {Carbajal}, \citenamefont {Martin},\ and\ \citenamefont
  {Chialvo}}]{Carbajal2022}%
  \BibitemOpen
  \bibfield  {author} {\bibinfo {author} {\bibfnamefont {J.~P.}\ \bibnamefont
  {Carbajal}}, \bibinfo {author} {\bibfnamefont {D.~A.}\ \bibnamefont
  {Martin}},\ and\ \bibinfo {author} {\bibfnamefont {D.~R.}\ \bibnamefont
  {Chialvo}},\ }\bibfield  {title} {\bibinfo {title} {Learning by mistakes in
  memristor networks},\ }\bibfield  {journal} {\bibinfo  {journal} {Physical
  Review E}\ }\textbf {\bibinfo {volume} {105}},\ \href
  {https://doi.org/10.1103/physreve.105.054306} {10.1103/physreve.105.054306}
  (\bibinfo {year} {2022})\BibitemShut {NoStop}%
\bibitem [{\citenamefont {Caravelli}(2017{\natexlab{b}})}]{Caravelli2016ml}%
  \BibitemOpen
  \bibfield  {author} {\bibinfo {author} {\bibfnamefont {F.}~\bibnamefont
  {Caravelli}},\ }\bibfield  {title} {\bibinfo {title} {The mise en sc{\'{e}}ne
  of memristive networks: effective memory, dynamics and learning},\ }\href
  {https://doi.org/10.1080/17445760.2017.1320796} {\bibfield  {journal}
  {\bibinfo  {journal} {Int. J. of Par., Em. and Dist. Sys.}\ }\textbf
  {\bibinfo {volume} {33}},\ \bibinfo {pages} {350} (\bibinfo {year}
  {2017}{\natexlab{b}})}\BibitemShut {NoStop}%
\bibitem [{\citenamefont {Bollob{\'{a}}s}(1998)}]{bollobas2012graph}%
  \BibitemOpen
  \bibfield  {author} {\bibinfo {author} {\bibfnamefont {B.}~\bibnamefont
  {Bollob{\'{a}}s}},\ }\href {https://doi.org/10.1007/978-1-4612-0619-4} {\emph
  {\bibinfo {title} {Modern Graph Theory}}}\ (\bibinfo  {publisher} {Springer
  New York},\ \bibinfo {year} {1998})\BibitemShut {NoStop}%
\bibitem [{\citenamefont {Sheldon}\ \emph {et~al.}(2020)\citenamefont
  {Sheldon}, \citenamefont {Caravelli},\ and\ \citenamefont
  {Coffrin}}]{coffrin}%
  \BibitemOpen
  \bibfield  {author} {\bibinfo {author} {\bibfnamefont {F.~C.}\ \bibnamefont
  {Sheldon}}, \bibinfo {author} {\bibfnamefont {F.}~\bibnamefont {Caravelli}},\
  and\ \bibinfo {author} {\bibfnamefont {C.}~\bibnamefont {Coffrin}},\
  }\bibfield  {title} {\bibinfo {title} {Fully analog memristive circuits for
  optimization tasks: a comparison},\ }\href@noop {} {\bibfield  {journal}
  {\bibinfo  {journal} {To appear in Handbook of Unconventional Computing, Ed.
  A. Adamatzky}\ } (\bibinfo {year} {2020})}\BibitemShut {NoStop}%
\bibitem [{\citenamefont {Pershin}\ \emph {et~al.}(2013)\citenamefont
  {Pershin}, \citenamefont {Slipko},\ and\ \citenamefont
  {Di~Ventra}}]{Pershin2013}%
  \BibitemOpen
  \bibfield  {author} {\bibinfo {author} {\bibfnamefont {Y.~V.}\ \bibnamefont
  {Pershin}}, \bibinfo {author} {\bibfnamefont {V.~A.}\ \bibnamefont
  {Slipko}},\ and\ \bibinfo {author} {\bibfnamefont {M.}~\bibnamefont
  {Di~Ventra}},\ }\bibfield  {title} {\bibinfo {title} {Complex dynamics and
  scale invariance of one-dimensional memristive networks},\ }\href
  {https://doi.org/10.1103/PhysRevE.87.022116} {\bibfield  {journal} {\bibinfo
  {journal} {Physical Review E}\ }\textbf {\bibinfo {volume} {87}},\ \bibinfo
  {pages} {022116} (\bibinfo {year} {2013})}\BibitemShut {NoStop}%
\bibitem [{\citenamefont {Sillin}\ \emph {et~al.}(2013)\citenamefont {Sillin},
  \citenamefont {Aguilera}, \citenamefont {Shieh}, \citenamefont {Avizienis},
  \citenamefont {Aono}, \citenamefont {Stieg},\ and\ \citenamefont
  {Gimzewski}}]{Sillin2013}%
  \BibitemOpen
  \bibfield  {author} {\bibinfo {author} {\bibfnamefont {H.~O.}\ \bibnamefont
  {Sillin}}, \bibinfo {author} {\bibfnamefont {R.}~\bibnamefont {Aguilera}},
  \bibinfo {author} {\bibfnamefont {H.-H.}\ \bibnamefont {Shieh}}, \bibinfo
  {author} {\bibfnamefont {A.~V.}\ \bibnamefont {Avizienis}}, \bibinfo {author}
  {\bibfnamefont {M.}~\bibnamefont {Aono}}, \bibinfo {author} {\bibfnamefont
  {A.~Z.}\ \bibnamefont {Stieg}},\ and\ \bibinfo {author} {\bibfnamefont
  {J.~K.}\ \bibnamefont {Gimzewski}},\ }\bibfield  {title} {\bibinfo {title} {A
  theoretical and experimental study of neuromorphic atomic switch networks for
  reservoir computing},\ }\href
  {https://doi.org/10.1088/0957-4484/24/38/384004} {\bibfield  {journal}
  {\bibinfo  {journal} {Nanotechnology}\ }\textbf {\bibinfo {volume} {24}},\
  \bibinfo {pages} {384004} (\bibinfo {year} {2013})}\BibitemShut {NoStop}%
\bibitem [{\citenamefont {Simmons}(1963)}]{Simmons1963}%
  \BibitemOpen
  \bibfield  {author} {\bibinfo {author} {\bibfnamefont {J.~G.}\ \bibnamefont
  {Simmons}},\ }\bibfield  {title} {\bibinfo {title} {Generalized formula for
  the electric tunnel effect between similar electrodes separated by a thin
  insulating film},\ }\href@noop {} {\bibfield  {journal} {\bibinfo  {journal}
  {Journal of Applied Physics}\ }\textbf {\bibinfo {volume} {34}},\ \bibinfo
  {pages} {1793} (\bibinfo {year} {1963})}\BibitemShut {NoStop}%
\bibitem [{\citenamefont {Bellew}\ \emph {et~al.}(2015)\citenamefont {Bellew},
  \citenamefont {Manning}, \citenamefont {{Gomes da Rocha}}, \citenamefont
  {Ferreira},\ and\ \citenamefont {Boland}}]{Bellew2015}%
  \BibitemOpen
  \bibfield  {author} {\bibinfo {author} {\bibfnamefont {A.~T.}\ \bibnamefont
  {Bellew}}, \bibinfo {author} {\bibfnamefont {H.~G.}\ \bibnamefont {Manning}},
  \bibinfo {author} {\bibfnamefont {C.}~\bibnamefont {{Gomes da Rocha}}},
  \bibinfo {author} {\bibfnamefont {M.~S.}\ \bibnamefont {Ferreira}},\ and\
  \bibinfo {author} {\bibfnamefont {J.~J.}\ \bibnamefont {Boland}},\ }\bibfield
   {title} {\bibinfo {title} {Resistance of {{Single Ag Nanowire Junctions}}
  and {{Their Role}} in the {{Conductivity}} of {{Nanowire Networks}}},\ }\href
  {https://doi.org/10.1021/acsnano.5b05469} {\bibfield  {journal} {\bibinfo
  {journal} {ACS Nano}\ }\textbf {\bibinfo {volume} {9}},\ \bibinfo {pages}
  {11422} (\bibinfo {year} {2015})}\BibitemShut {NoStop}%
\end{thebibliography}%

\clearpage
\appendix

\section{Memristive models}
\subsection*{$TiO_2$ toy model} 
\label{sec:tmod}

For a circuit with multiple memristors having varying resistances $R(x_i)$ and constant voltage generators $S_i=\tilde{S}$ connected in series, the equation (\ref{eq:oned}) for a single memristor extends to a system of coupled nonlinear ordinary differential equations. This system describes the network dynamics of the memory elements $x_i(t)$ and can be expressed as \cite{Caravelli2016rl,Caravelli2019Ent}:
\begin{eqnarray}
\frac{d}{dt} \vec x=\frac{1}{\beta}(I-\chi {\mathcal P} X)^{-1} {\mathcal P} \vec S-\alpha \vec x,
\label{eq:manyd}
\end{eqnarray}
Here, $\chi=\frac{R_{off}-R_{on}}{R_{off}}<1$ and $X_{ij}(t)=x_i(t) \delta_{ij}$. The matrix ${\mathcal P}$ represents the projection operator on the vector space of cycles in the circuit's graph $\mathcal{G}$ \cite{Caravelli2016rl}, and it can be obtained from the directed incidence matrix $B$ of $\mathcal{G}$ as ${\mathcal P}=I-B(B^t B)^{-1} B^t$ \cite{Caravelli2017b,Caravelli2016ml,Nilsson,bollobas2012graph}. The property of ${\mathcal P}$ being a projection operator reflects Kirchhoff's circuit laws. In this paper, we consider $B$ as a random matrix to abstract the dynamical system from any specific circuit topology.

It is interesting to see that there is a range of voltage in which the system experiences bistability. This is not  a pitchfork bifurcation, but it is nonetheless the emergence of two (meta)stable points in the range $[Vc,Vc^*]$ of Fig. \ref{fig:mfp} (bottom). We can see the relevance of this feature in the study of ergodicity as a function of voltage, as we explain below.

In order to study the ergodic vs. non-ergodic behavior of such toy model, we extend the system to a stochastic one, introducing noise.
The dynamical equations of a memristive device can incorporate noise through stochastic terms, introducing randomness into the system's behavior. This stochasticity is manifested in the memory component of the device. This is a diffusive memristive model of the form
\begin{align}
  d \vec x 
&=  
    \frac{1}{\beta}
    \left[
        I
    -
        \chi
        {\mathcal P}
        X
    \right]^{-1}
    {\mathcal P}
    \vec S
    dt
- 
    \alpha 
    \vec x 
    dt
    \nonumber
    \\
&\ +
    \sigma 
    (\vec x, t)d
    \vec \eta
    ,
\label{eq:manydnoise}
\end{align}
%
the vector $d \vec \eta$ contains $n$ Wiener stochastic processes  with $\sigma$ the diffusion coefficient.


It is known that for the memristive toy model of eqn. (\ref{eq:manyd}) and in the limit of dense networks, for values of $\chi=0.9$ and $\bar s=\frac{1}{\alpha \beta} \frac{1}{N}\sum_{i=1}^N S_i\approx 0.24$ in which the system develops two mean-field minima, induces bulk dynamical transition characterized by a short transient of positive Lyapunov exponents \cite{caravelli2021}. For smaller or larger values of $\bar s$ instead, the memory parameters are in a laminar regime, characterized by the presence of only a global minimum in the potential $V(x,s)$. 

We discuss the properties of the effective potential.
For $\alpha=\beta=1$ and $\chi=0.9$, the range is $1/10< s <5/18$. The emergence of the two minima is non-perturbative requiring $\chi$ to be close to 1. In the case of $\chi=0$, the network consists of regular resistors, and the behavior of the single-element dynamical system is characterized by simple basins of attraction without any exotic features. To ensure the validity of the equations of motion for the single variables, cutoff functions, such as $\frac{d}{d\tau} x=-W(x) f(x,y)$, are applied, where $W(x)=1$ for ${0< x< 1} \cup{x=1,f(x,y)>0}\cup{x=0,f(x,y)<0} $, and zero otherwise. An implicit analytical solution for this differential equation has been derived in \cite{coffrin}.



\subsection*{Junction Model}\label{sec:juncm}


Nanowire-nanowire cross-points are modeled as electrically insulating, ionically conducting junctions \cite{Pershin2013, Sillin2013, Diaz-Alvarez2019, Kuncic2020}. The junctions exhibit voltage-threshold memristive switching due to electro-chemical metallization and the influence of electron tunnelling transport \cite{Hochstetter2021}. However, the model neglects junction-level stochastic fluctuations, which introduces further nonlinearities and lead to more complex dynamics \cite{Hochstetter2021}.
Junction conductance, $G = G(\lambda)$, is determined by a state variable $\lambda(t)$ which characterizes the length of the conducting filament responsible for memristive switching:

\begin{align}
        G(\lambda) = \frac{1}{R_\mathrm{t}(\lambda) + R_\mathrm{on} + \xi} + 
        \frac{1}{R_\mathrm{off}},
\end{align}
where 

\begin{align}
    \xi = \frac{R_\mathrm{on}^2}{R_\mathrm{off} - R_\mathrm{on}},
\end{align}
and the tunnelling conductance $G_t(\lambda)$ is calculated using the low voltage Simmon's formula (Eq.~\ref{eq:tunSwitchCon}) for a MIM junction \cite{Simmons1963}:

\begin{eqnarray}
        G_t (\lambda)&=& [R_\mathrm{t}(\lambda)]^{-1} \nonumber \\
        &=& \frac{ 3(2m_*)^{1/2}e^{5/2}   (\phi/e)^{1/2}}{A 2h^2 s^2} \nonumber \\
        & &\hspace{0.1cm}\cdot
        \exp \left[-\frac{4\pi (2m_* e)^{1/2}}{h}  s \left(\frac{\phi}{e}\right)^{1/2} \right]
    \label{eq:tunSwitchCon}\\
    s &=& \max\left[\left(\Lambdac-|\Lambda|\right)\frac{\smax}{\Lambdac} \; , \; 0\right],  \label{eq:lambdaS}
\end{eqnarray}
with effective mass $m_*=0.99 \me$ and PVP layer (assumed homogeneous) thickness $\smax = 5\,\text{nm}$. 
The difference between Fermi levels of PVP and Ag determines the potential barrier $\phi = 0.82\,\text{eV}$.
$A = (0.41\,\text{nm})^2  = 0.17 \,\text{nm}^2$ is the area of a face of the silver unit cell. 
The resistances of nanowires are considered to be negligible compared to the resistance of the junctions \cite{Bellew2015}. Consequently, $G_t(\lambda)$ exhibits an additional nonlinear dependence on $V$, through the filament growth parameter $s = s(\Lambda(V))$ which modulates junction switching due to filament formation.

All junctions start with a high resistance ``off'' state initially. When the value of $\lambda$ of a single junction reaches the predefined threshold $\lambda_{\text{crit}}$, the resistance of the junction switches to ``on'' state.
The ratio of these resistance states is $R_{\text{off}}/ R_{\text{on}} = 10^3$,
with $R_{\text{on}} = 10^4 $ Ohms.

The evolution of $\lambda(t)$ is described by a polarity-dependent voltage-threshold model \cite{Pershin2013, Kuncic2020,Zhu2021information, Hochstetter2021}:
\begin{equation}
    \dfrac{d \lambda}{dt} = 
        \begin{cases}
            (|V(t)| - V_{\text{set}}) \text{sgn}[V(t)], & |V(t)| > V_{\text{set}}\\
            0 , & V_{\text{reset}} < |V(t)| < V_{\text{set}}\\
            b(|V(t)| - V_{\text{reset}}) \text{sgn}[\lambda(t)], & |V(t)| < V_{\text{reset}}
        \end{cases}
\end{equation}
where $V_{\text{set}}$ is the on-threshold and $V_{\text{reset}}$ is the off-threshold, and $b$ is a positive constant defining the relative rates of decay of the filament. 
All simulation results presented in this work are generated using the following parameters:
$V_{\text{set}} = 10^{-2}$V, $V_{\text{reset}} = 10^{-3}$V and $b = 10$.
Experimental validation of this model is presented elsewhere \cite{Diaz-Alvarez2019, Hochstetter2021}.

More specifically, the junction voltage distribution $V$ across the network is obtained numerically by solving 
\begin{equation}
{\cal L}^\dagger V = I \qquad ,
\end{equation}
in which $I$ is current and $\cal{L}^\dagger$ is the expanded graph Laplacian of the network, expressed in block matrix representation as 
\begin{equation}
	\cal L^\dagger = 
	\left[
	\begin{array}{c|c}
	\cal L & C\\ 
	\hline
	C & 0\\
	\end{array}
	\right] \qquad ,
\end{equation}
where $\cal L$ is the graph Laplacian and where the elements of $C$ are either 1 if the nanowire node is connected to an external electrode or 0 otherwise. 
The Laplacian $\cal L$ is generated by
\begin{equation}
{\cal L} = D - W \qquad ,
\end{equation}
where $W$ is the weighted adjacency matrix of the network, with the weights representing junction conductance distribution:
\begin{equation}
    W_{ij} = A_{ij} G(i,j) \qquad ,
\end{equation}
where $G(i,j)$ is conductance on the edge connecting nodes $i$ and $j$ and $D$ is the weighted degree matrix generated from $W$:
\begin{equation}
    D = \textbf{diag}(d_i) \qquad , \qquad d_i = \sum \limits_{k=1}^{N} W_{ik} \qquad .
\end{equation}
After Kirchoff laws are solved, the parameters $\lambda(t)$ are evolved accordingly.



\section{Reservoir computing and multiplexing}\label{app:b}
Reservoir computing is a popular approach to machine learning that utilizes a fixed, randomly connected network of nodes, known as a reservoir, to generate temporal patterns in response to an input signal. The reservoir acts as a dynamical system that maps input signals to high-dimensional feature spaces, where a readout layer can then be trained to classify or predict the desired output. This approach is particularly effective for processing time-varying data and has been applied to a range of tasks, such as speech recognition, natural language processing, and control systems. One of the key advantages of reservoir computing is that the reservoir can be pre-designed and does not require tuning during the training phase, reducing the complexity of the learning process. Furthermore, the reservoir can be implemented using a variety of physical substrates, such as electronic circuits, optical systems, and biological neural networks, making it a versatile approach that can be adapted to a wide range of applications.

The training of a reservoir computer involves finding a set of optimal readout weights that map the reservoir states to the desired output. This can be formulated as a linear regression problem, where the readout weights $\mathbf{W}_{out}$ are estimated by minimizing the mean squared error between the target output $\mathbf{y}(t)$ and the estimated output $\hat{\mathbf{y}}(t)$:

\begin{equation*}
\min_{\mathbf{W}{out}} \frac{1}{T} \sum_{i=1}^{T} |\mathbf{y}(t_i) - \mathbf{W}_{out} \mathbf{x}(t_i)|^2
\end{equation*}

where $\mathbf{x}(t_i)$ is the reservoir state at time step $n$, and $N$ is the total number of training samples. The solution to this optimization problem can be obtained using the pseudo-inverse of the reservoir state matrix $\mathbf{X}$:

\begin{equation*}
\mathbf{W}_{out} = (\mathbf{X}^t \mathbf{X} + \alpha \mathbf{I})^{-1} \mathbf{X}^t \mathbf{Y}
\end{equation*}
with $\ ^t$ representing the vector or matrix transposition,
where $\mathbf{Y}$ is the target output matrix, and $\alpha$ is a regularization parameter that prevents overfitting. The reservoir state matrix is defined as:

\begin{equation*}
\mathbf{X} = [\mathbf{x}(1), \mathbf{x}(2), ..., \mathbf{x}(N)]^t
\end{equation*}
and the target output matrix is defined as:

\begin{equation*}
\mathbf{Y} = [\mathbf{y}(1), \mathbf{y}(2), ..., \mathbf{y}(N)]^t
\end{equation*}

Once the readout weights are estimated, they can be used to generate the output for new input signals by computing:

\begin{equation*}
\hat{\mathbf{y}}(t_i) = \mathbf{W}_{out} \mathbf{x}(t_i)
\label{eq:train}
\end{equation*}

where $\hat{\mathbf{y}}(t_i)$ is the estimated output at time step $t$. In the following, we will assume that $\mathbf{x}(t)$ are either the internal memory states or the conductance states.

In the case of a single conductance measured, in order to multiply the number of virtual nodes, we used the multiplexing technique introduced in \cite{Appeltant2011}. The time series is time multiplexed at certain sampling frequencies. Then, the time series is parsed into $N$ vectors of length $K$, such that $NK$ is the length of the time series. Then, elements of such vectors at locations $mN+l$ for integer $l\in[1,K]$ are considered as different internal virtual nodes. The regression then follows the same scheme as the previous section. 

Reservoir computing has been successfully applied in various physical systems, including opto-electronic and nanoscale systems such as silver nanowire networks (NWNs) to implement neuromorphic computing. NWNs are particularly interesting for this application due to their memristive behavior and complex network topology that resembles that of biological neural networks. By leveraging the dynamics of NWNs as a reservoir, the input signals can be mapped to a high-dimensional state space through a random projection, followed by a simple readout operation to obtain the desired output. This allows for the implementation of complex computational tasks such as pattern recognition, time-series prediction, and control in a highly efficient and parallel manner. In this way, reservoir computing offers a promising approach to harness the capabilities of NWNs for neuromorphic computing, and silver nanowire learning represents an exciting area of research at the intersection of nanotechnology and machine learning.

To be considered a feasible reservoir, a dynamical system must satisfy the condition that its state approaches a nontrivial function of the input trajectory in the long time limit. This can be expressed in terms of two requirements\cite{sheldonrc}:

\emph{Fading Memory}: If the system were started from two different initial conditions while driven with the same input trajectory $u$, the system's trajectories should eventually converge to the same state. This implies that the system has a finite temporal memory;

\emph{State Separation}: Different input sequences should drive the system into different trajectories. That is, if the same initial condition were driven with two different input trajectories, the resulting reservoir trajectories must be sufficiently different. 
These properties could be proven analytically for the case of the memristive reservoir toy model.

These conditions ensure that the state of the reservoir becomes a function of the input trajectory and that this function carries information about the input trajectory. In the case of memristive systems, the memristive junctions act as the dynamic elements of the reservoir, and the network dynamics arise from the collective behavior of the junctions. The input signal drives the memristive junctions to switch between high and low resistance states, resulting in a temporal sequence of network states. The final state of the network can be used as a high-dimensional feature vector for further processing. In the case of silver nanowire networks, the nanowires themselves act as the memristive elements, and their collective behavior gives rise to the network dynamics. The input signal drives the nanowires to switch between metallic and insulating states, resulting in a temporal sequence of network states that can be used for further processing.

As we saw above, reservoir computing requires the fading memory property for the dynamics of a reservoir, which is equivalent to the system tending to some degree of effective ergodicity when expressed in physical terms. The scope of this manuscript is to investigate fading memory and understand how quickly a reservoir ``forgets" its initial conditions, using the Thirumalai-Mountain metric as our benchmark. The goal is to determine the optimal level of memory capacity for different system parameters.
However, there are limitations to using ergodicity and the ergodic theorem, as ensemble averages are only defined for systems at equilibrium. Nonetheless, we can still consider the case of systems locally at equilibrium.

The original formulation of the TM metric aimed to predict the dynamics of a system numerically simulated on a computer without having to calculate the Lyapunov exponents of systems with a high number of variables. This is particularly relevant when dealing with large-scale systems, where calculating Lyapunov exponents can be computationally expensive. 


Let us now briefly introduce the time series we used for our benchmarks.
The Mackey-Glass time series is a widely used benchmark for testing the performance of time-series prediction models. It is a nonlinear, nonautonomous, and nonperiodic time series that exhibits chaotic dynamics. The series is defined by the following difference equation:

\begin{equation}
\frac{dx(t)}{dt} = r \frac{x(t-\tau)}{1+x(t-\tau)^n} - \gamma x(t),
\end{equation}

where $x(t)$ is the value of the time series at time $t$, $\beta$, $\gamma$, $n$, and $\tau$ are parameters that control the dynamics of the system. The Mackey-Glass time series exhibits a range of dynamic behaviors, depending on the values of these parameters. For example, for $r= 0.2$, $\gamma = 0.1$, $\tau = 17$, and $n=10$, the time series exhibits chaotic dynamics with a positive Lyapunov exponent.



\end{document}